\documentclass[a4paper,11pt]{article}

\usepackage{natbib}
\usepackage{eqnarray,amsmath,amsthm}
\usepackage{amssymb}
\usepackage{graphicx}
\usepackage{authblk} 
\usepackage{mathtools}
\usepackage{mathabx}
\usepackage{xcolor,colortbl}
\usepackage{subfigure}
\usepackage[margin=3cm]{geometry}
\usepackage{float}
\usepackage{nicefrac}
\usepackage{hyperref}
\usepackage{microtype} % better spacing and less overfull \hbox warnings
\usepackage{enumitem}
\usepackage{nicematrix}
\usepackage{bbm}
\usepackage{hyperref}
\usepackage{changepage}
\usepackage{algorithm}%
\usepackage{algorithmicx}%
\usepackage{algpseudocode}%
\usepackage{tikz}
\usetikzlibrary{positioning, fit, backgrounds, arrows.meta}

\makeatletter
\algnewcommand{\LineComment}[1]{{\Statex \hskip\ALG@thistlm \footnotesize\textcolor{blue}{/* #1 */}}}
\algnewcommand{\LineCommentIndent}[1]{{\Statex \hskip\ALG@tlm \footnotesize\textcolor{blue}{/* #1 */}}}
\makeatother

\newcommand{\Cat}{\mathrm{Cat}}

\newcommand{\proptos}{\ensuremath{\, \propto \,}}

\title{A Bayesian spatio-temporal nearest neighbor Gaussian process model for pooled genetic data}

\author[1]{Imke Botha} 
\author[2,3]{Tianxiao Hao}
\author[4,5,6]{Lucinda E. Harrison}
\author[2,7]{Nick Golding}
\author[2,7]{Daniel J. Weiss}
\author[1]{Jennifer A. Flegg}

\affil[1]{School of Mathematics and Statistics, University of Melbourne, Australia}
\affil[2]{The Kids Research Institute of Australia, Australia}
\affil[3]{Melbourne School of Population \& Global Health, University of Melbourne, Australia}
\affil[7]{School of Physics, Mathematics and Computing, The University of Western Australia}
\affil[4]{WorldWide Antimalarial Resistance Network (WWARN), Oxford, UK}
\affil[5]{Infectious Diseases Data Observatory (IDDO), Oxford, UK}
\affil[6]{Centre for Tropical Medicine and Global Health, Nuffield Department of Medicine, University of Oxford, UK}

\begin{document}
	\setlength{\parindent}{0pc}
	\setlength{\parskip}{1ex}
	\maketitle
	\begin{abstract}
		Large scale genetic datasets often aggregate the total allele counts of distinct genetic markers. Inferring haplotype frequencies (i.e.\ the frequency of multimarker alleles) from these pooled data is a challenge. Previous spatio-temporal modelling in this context has been limited to 3 markers due to the computational cost. In this work, we propose a nearest neighbor Gaussian process (NNGP) model to improve scaling with the number of markers and observations. To infer the parameters of our model, we develop a novel sequential Monte Carlo squared algorithm, which uses particle Gibbs with ancestor sampling to mutate the NNGP function values. The latter has a linear cost in the number of observations and the number of NNGPs, and can be applied to a broad range of NNGP models. As a case study, we analyse genetic data relating to antimalarial drug resistance in Africa, and show our scaling results empirically on a 3 and 6 genetic marker dataset. 
	\end{abstract}


	\section{Introduction}

    Large-scale genetic studies often genotype combined DNA from groups of individuals as a cost-effective alternative to genotyping each DNA sample separately. Genotyping is performed at specific genetic markers, defined here as single nucleotide polymorphisms (SNPs), which are positions on a chromosome where a single nucleotide can vary between individuals. Counts of each possible variant at the SNP (alleles) are reported. 
    
    Pooled genetic data are often used for genetic association, which aims to link alleles to specific traits, such as drug resistance, or diseases within a population. Often, it is the particular combination of marker alleles on the same chromosome, called a haplotype, that is of interest \citep{wright2005,tam2019,vaneijk2019}. These data may subsequently be used to track the spread of alleles or haplotypes associated with those traits or diseases.

    Inferring haplotype prevalences is challenging, as pooled genetic data generally report aggregated allele counts. When markers of interest lie on different chromosomes, only single-marker allele counts are generally reported for each pool. In this case, the data do not identify which alleles co-occur within the same individual genome. Even for markers that lie on the same chromosome, a study may still report aggregated counts either because of the genotyping method used or to simplify reporting. 

	Several methods to infer haplotype prevalences from pooled data have been developed. A fairly general approach is to model the unobserved haplotype counts as latent variables following a multinomial distribution \citep{ito2003,iliadis2012,foo2023}. The inferred counts must be consistent with the observed data, e.g.\ the sum of haplotype counts involving a specific marker allele must equal the observed count for that marker allele. To ensure this, some methods enumerate all feasible sets of haplotype counts \citep{iliadis2012,foo2023}, however, these methods do not scale well with pool size, i.e.\ the number of DNA samples per pool. \citet{iliadis2012} use a Bayesian tree-based approach to estimate haplotype frequencies for each feasible set, while \citet{foo2023} calculate the likelihood by summing over the feasible sets, and use Hamiltonian Monte Carlo for parameter inference. \citet{foo2023} also propose a Metropolis-within-Gibbs sampling method, where the sampled latent counts are consistent with the observations by construction, i.e.\ without needing to enumerate feasible haplotype counts. The latter approach scales linearly with pool size, but is still limited to a small number of markers. 
		
	A different strategy is to use a multivariate normal approximation of the multinomial distribution, which approximately integrates over the latent counts. \citet{kuk2009} use the normal approximation and propose an approximate expectation-maximisation method for estimation of the haplotype frequencies. \citet{pirinen2009} extends this method by allowing estimation of a smaller subset of input haplotypes, prespecified by the user. They also introduce a second approach which uses reversible jump Markov chain Monte Carlo (MCMC) to estimate the list of input haplotypes alongside the haplotype frequencies. While the normal approximation scales very well computationally with pool size and the number of markers, some haplotype frequencies (those near 0 or 1) can lead to singular or near-singular covariance matrices. While additional approximations can ensure the covariance matrix is non-singular, e.g.\ by adding a small positive constant to the diagonals, near-singularity can still lead to a poor approximation \citep{foo2023}. 
	
	Our interest is specifically in inferring haplotype frequencies from pooled genetic data in the spatio-temporal setting, i.e.\ where data are collected from multiple studies over different years and locations. \citet{foo2023} introduce a hierarchical model, where the frequencies of each haplotype is modelled using a spatio-temporal Gaussian Process (GP). The GP allows for spatial and temporal variation in the inferred haplotype frequencies, but adds further computational constraints. The cost of the GPs in this setting is $O((H-1)D^3)$, where $H$ is the number of haplotypes, and $D$ is the number of pools. 
	
    We adapt the model of \citet{foo2023} to improve scalability with respect to the number of markers and pools. In particular, we use a GP for each haplotype as in \citet{foo2023} combined with a Beta-Binomial composite marginal likelihood. This choice of likelihood assumes the observations are independent given the haplotype prevalences, but does not require the haplotype counts to be inferred. To also reduce the cost of the GPs, we use nearest neighbor GPs (NNGPs) \citep{datta2016}, an approach which has complexity of order $O((H-1)Dk^3)$, where $k$ is fixed and typically much smaller than $D$. 

	A sequential Monte Carlo \citep[SMC; ][]{delmoral2006} algorithm is used to infer the model parameters. The SMC mutation kernel is a Metropolis-within-Gibbs algorithm, where the NNGP function values are updated using particle Gibbs with ancestor sampling \citep[PGAS; ][]{lindsten2014}. We show how PGAS can be applied in this setting such that the cost of the algorithm is linear with respect to the number of pools and the number of haplotypes. It is straightforward to apply this method to other models as well and it can be seen as a general approach for Bayesian inference of NNGP models. Our proposed method yields exact parameter inference in the pseudo marginal sense \citep{andrieu2009}, and falls within the SMC squared framework \citep[SMC$^2$; ][]{chopin2012,duan2014}.
	
	As a case study, we apply our approach to pooled genetic data relating to antimalarial drug resistance in Sub-Saharan Africa. The data were compiled through a systematic review, with each pool generally corresponding to a different, independently administered study. Consequently, there is substantial between-pool heterogeneity, which is partly reflected in differences in pool size and the reported markers and haplotypes \citep{iddo2026}.
    
    The rest of the paper is organized as follows. Section \ref{sec:model} describes the spatio-temporal NNGP model. Section \ref{sec:methods} describes SMC$^2$ and the proposed PGAS approach. Section \ref{sec:example} shows the application of our method to the antimalarial drug resistance case study for 3 and 6 markers and Section \ref{sec:discussion} concludes with a discussion.

    \section{Hierarchical NNGP Model} \label{sec:model}

    \subsection{Spatio-temporal Pooled Genetic Data}
    
    Our aim is to infer haplotype prevalences based on $G$ biallelic markers, i.e.\ where there are two possible alleles for each marker. Haplotypes are represented by a binary string of length $G$, where the baseline (or wildtype) allele is represented by $0$ and the other allele (referred to as the mutation) is represented by $1$. There are $H = 2^G$ possible haplotypes for a set of $G$ markers.
    
    We assume data are available from $D$ studies spanning multiple years and locations. Each study considers a subset of the $G$ markers of interest, and reports counts of each mutation. If genotyping is done on a single chromosome, the study may also report on the co-occurrence of any subset of the marker alleles. For study $d$, $d = 1,\ldots, D$, let $y_d$ be the vector of $h_d \le H$ observed counts, $n_d$ be the vector of the total number of samples tested for each observed count in $y_d$, and $x_d$ be the vector of spatio-temporal covariates. Note that the reported sample size may differ between counts within a single study due to marker-specific measurement error or post-processing quality control \citep{sham2002,barratt2002}. 
    
    Each study also has a vector of $H$ latent or unobserved counts $z_d$, and a configuration matrix $A_d \in \{0,1\}^{h_d \times H}$ that encodes which haplotypes the study has information on. Observations are recovered by $y_d = A_d z_d$. 
    
    \textbf{Example} \\
    As an example, if there are $G=3$ markers of interest (giving $H=8$ possible haplotypes), the $d$th vector of latent counts is $z_d = (z_{000},z_{100}, z_{010}, z_{001}, z_{110}, z_{101}, z_{011}, z_{111})^{\top}$. Denote the absence of information about an allele by $-1$. A typical study might only report on a subset of the markers, e.g.\ $y_d = (y_{(1,-1,-1)}, y_{(-1,1,-1)}, y_{(1,1,-1)})^{\top}$, where 
    \begin{align*}
    	y_{1,-1,-1} &= z_{100} + z_{110} + z_{101} + z_{111} \\
    	y_{-1,1,-1} &= z_{010} + z_{110} + z_{011} + z_{111} \\
    	y_{1,1,-1} &= z_{110} + z_{111}.
    \end{align*}
    The configuration matrix for this study is
    $$
    	A_d = 
    	\begin{pNiceMatrix}[first-row,first-col]
			    	  & 000 & 100 & 010 & 110 & 001 & 101 & 011 & 111 \\
    		(\phantom{-}1,-1,-1) & 0 & 1 & 0 & 1 & 0 & 1 &	0 & 1  \\
    		(-1,\phantom{-}1,-1) & 0 & 0 & 1 & 1 & 0 & 0 & 1 & 1  \\
    		(\phantom{-}1,\phantom{-}1,-1)  & 0 & 0 & 0 & 1 & 0 & 0 & 0 & 1
    	\end{pNiceMatrix}.
    $$
   In the rest of the paper, we refer to the observed marker or multimarker mutations, e.g.\ $(1,-1,-1)$, $(-1,1,-1)$ and $(1,1,-1)$, as the observed haplotypes.
	
	\subsection{Nearest Neighbor Gaussian Process} \label{ssec:nngp}
	
	To improve scaling with the number of observations, a sparse GP can be used. A non-degenerate approach is the nearest neighbor GP (NNGP), which is based on Vecchia's approximation \citep{vecchia1988}. See \citet{datta2016} for more details. 
	
	The full GP for each haplotype $h$ can be written as $\mu_h(x_d) + f_h(x_d)$, where $\mu_h(x_d)$ is the mean function and
	\begin{align*}
		f_h(x_d) &\sim \mathcal{GP}  (k_h(x_d, x_{d'})) \\
		f_{h,1:D} &= f_h(x_{1:D}) = (f_h(x_1), \ldots, f_h(x_D))^{\top} \sim \mathcal{N}(0, C_h),
	\end{align*}
	is the zero-mean GP component. The $(a, b)$th element of $C_h$ is given by $k(x_{a}, x_{b})$. The density of $f_{h,1:D}$ can be recursively written as
	\begin{align*}
		p(f_{h, 1:D}) &= p(f_{h, 1})\prod_{d = 2}^{D}p(f_{h, d}\mid f_{h, 1:d-1}).
	\end{align*}
	The NNGP is constructed from the full GP based on neighbor sets associated with each location $x_d$. Let $K(x_{d})$ be the set of indices of the (at most) $k$ nearest neighbors of $x_{d}$ in $\{x_1, \ldots, x_{d-1}\}$. The density of $f_{h,1:D}$ then becomes
	\begin{align*}
		p(f_{h, 1:D}) = p(f_{h, 1})\prod_{d = 2}^{D}p(f_{h, d}\mid f_{h, K(x_{d})}).
	\end{align*}
	The NNGP is 
	\begin{align*}
		f_h(x_d) &\sim \mathcal{NNGP}(k_h(x_d, x_{d'})) \\
		f_{h,1:D} &\sim \mathcal{N}(0, \tilde{C}_h), \quad
		\tilde{C}_h = \left((I-A_h)^{\top}{D_h}^{-1}(I-A_h)\right)^{-1},
	\end{align*}
	where $A_h$ is a sparse lower triangular matrix with at most $k$ non-zero entries in each row, and $D_h$ is a diagonal matrix \citep{finley2017}. The non-zero entries of $A_h$ and $D_h$ are
	\begin{align*}
		A_h(d, K(x_d)) &= C_h(x_d, K(x_d))(C(K(x_d), K(x_d)))^{-1} \\
		D_h(d,d) &= C_h(x_d, x_d) - {}\\
		&\qquad C_h(x_d, K(x_d))(C_h(K(x_d), K(x_d)))^{-1}C_h(K(x_d), x_d).
	\end{align*}
	The predictive distribution of $f_{h, d}$ given the neighbors simplifies to 
	\begin{align}
		p(f_{h, d}\mid f_{h, K(x_{d})}) &= \mathcal{N}(f_h(x_d)\mid A_h(d, K(x_d))f_{h, K(x_d)}, D_h(d,d)). \label{eqn:nngp_pred}
	\end{align} 
	
	\citet{datta2016} find that the ordering of the points do not generally have a significant impact on the results. A sensible ordering for spatio-temporal molecular marker data is to order by time. This follows \citet{datta2016a}, and ensures that neighbors cannot be from `future' time points. \citet{datta2016} also find that a modest number of neighbors ($k<20$) is sufficient in most settings. In all of our experiments in Section \ref{sec:example}, $D\le 300$ and we use a conservative $k=20$ neighbors.

	\subsection{Spatio-temporal Nearest Neighbor GP Model}
	
	The haplotype prevalences are given by a softmax transformation of the $H$ NNGPs,
	\begin{align*}
		p_{h, d} &= \frac{\exp{(\mu_h(\epsilon_h, x_d) + f_h(x_d))}}{\sum_{h = 1}^{H}\exp{(\mu_h(\epsilon_h, x_d) + f_h(x_d))}} \\
		p(f_{h, 1:D}\mid\eta_h) &\proptos \mathcal{N}(0, \tilde{C}_h(\eta_h)),
	\end{align*}
	where $\mu_h(\epsilon, x_d)$ is the mean function and $\epsilon_h$ and $\eta_h$ are vectors of parameters for the $h$th NNGP. Since the softmax transformation is translation invariant, only $H-1$ of the NNGP function values $f_{1:H, d}$ are identifiable. To improve identifiability, a sum-to-zero constraint is applied using a contrast matrix reparameterisation \citep{wood2017}, where the contrast matrix is chosen to be the orthonormal Helmert submatrix, denoted by $B$ \citep{anderson2003}. This restricts $f_{d}$ to a $J=H-1$ dimensional subspace, such that
	\begin{align*}
		p_{h, d} = \frac{\exp{(\tilde{f}_{h, d})}}{\sum_{m = 1}^{H}\exp{(\tilde{f}_{m, d})}},
	\end{align*}
	where $\tilde{f}_{1:H, d} = B(\mu_{1:J}(\epsilon_{1:J}, x_d) + f_{1:J}(x_d))$.
	
	A hierarchical prior is given to $\eta_{1:J}$, $p(\eta_{1:J},\phi) = \prod_{j = 1}^{J}p(\eta_j\mid\phi)p(\phi)$. The density of the NNGP function values is
	\begin{align}
		f_{1:J, d} = (f_1(x_d),\ldots, f_J(x_d))^{\top} &\sim p(f_{1:J,d}\mid f_{1:J, K(x_d)}, \eta_{1:J}) \nonumber \\
		&= \mathcal{N}(m_d(f_{1:J, K(x_d)}, \eta_{1:J}), C_{d}(\eta_{1:J})), \label{eqn:transition}
	\end{align}
	where $m_d(f_{1:J, K(x_d)}, \eta_{1:J})$ is a vector of the same length as $K(x_d)$. The $j$th entry of the mean vector $m_d(f_{1:J, K(x_d)}, \eta_{1:J})$ is given by $A_j(d, K(x_d))f_{j, K(x_d)}$, and $C_{d}(\eta_{1:J})$ is a diagonal matrix with $j$th diagonal entry given by $D_j(d,d)$; see Section \ref{ssec:nngp}. The density of $f_{1:J, 1:D}$ is
	\begin{align*}
		p(f_{1:J, 1:D}\mid\eta_{1:J}) = \prod_{j = 1}^{J}p(f_{j, 1:D}\mid\eta_j).
	\end{align*}
	
	Given the vector of $H$ haplotype prevalences $p_d$ and $H$ latent counts $z_d$, the likelihood is 
	\begin{align*}
		p(z_{d} \mid f_{1:J, d}, \epsilon_{1:J}) = \operatorname{Multinomial}(n_{d}, p_{d}), \quad y_d = A_d z_d
	\end{align*}
	or, in the case of overdispersion, e.g.\ due to variation between studies,
	\begin{align*}
		p(z_{d} \mid f_{1:J, d}, \epsilon_{1:J}, \kappa) = \operatorname{DirichletMultinomial}(n_{d}, \kappa p_{d}), \quad y_d = A_d z_d
	\end{align*}
	where $\kappa$ is the precision parameter. To avoid the problem of inferring $z_d$ subject to the constraint $y_d = A_d z_d$, and to allow for different sample sizes within the same study, we define $\tilde{p}_d = A_{d}p_{d}$ as the vector of observed prevalences and instead use a composite marginal likelihood \citep{varin2008,varin2011}, 
	\begin{align*}
		p(y_{d} \mid f_{1:J, d}, \epsilon_{1:J}) \proptos \prod_{i = 1}^{h_d} \operatorname{Binomial}(y_{i,d} \mid n_{i,d}, \tilde{p}_{i,d})^{w_{i,d}},
	\end{align*}
	or
	\begin{align}
		p(y_{d} \mid f_{1:J, d}, \epsilon_{1:J}, \kappa) \proptos \prod_{i = 1}^{h_d} \operatorname{BetaBinomial}(y_{i,d} \mid n_{i,d}, \kappa\tilde{p}_{i,d}, \kappa (1-\tilde{p}_{i,d}))^{w_{i,d}}, \label{eqn:likelihood}
	\end{align}
	where $y_{i,d}$ is the $i$th element of $y_d$, $n_{i,d}$ is the $i$th element of $n_d$, $\tilde{p}_{i,d}$ is the $i$th element of $\tilde{p}_{d}$ and $w_{i,d}$ is the $i$th element of the composite weights $w_d$. The simplest choice is to use unit weights, i.e.\ $w_{i,d} = 1$, but the weights can also be chosen to reduce redundancy among overlapping likelihood components \citep{varin2011,bevilacqua2015}. See Section \ref{sec:example} for more details.
	
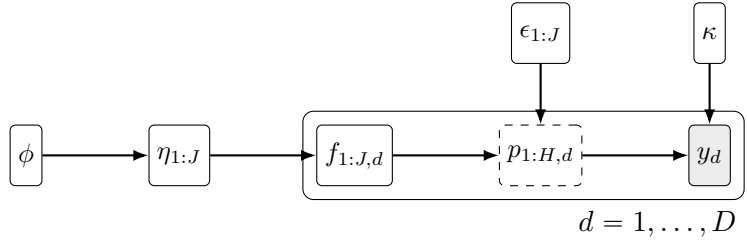
\begin{figure}[ht]
		\centering
		\begin{tikzpicture}[
			node distance = 10mm and 14mm,
			latent/.style = {
				draw,
				rounded corners=2pt,
				minimum height=8mm,
				inner sep=3pt,
				align=center,
				font=\small},
			obs/.style = {
				latent,
				fill=gray!15},
			det/.style = {
				latent,
				dashed},
			arr/.style = {-{Latex[length=2mm]}, thick},
			plate/.style = {
				draw,
				rounded corners,
				inner sep=5pt,
				font=\small}]
			
			% nodes
			\node[latent] (phi) {$\phi$};
			\node[latent, right=of phi] (eta) {$\eta_{1:J}$};
			\node[latent, right=of eta] (f) {$f_{1:J,d}$};
			\node[det, right=of f] (p) {$p_{1:H,d}$};
			\node[obs, right=of p] (y) {$y_d$};
			\node[latent, above=8mm of p] (eps) {$\epsilon_{1:J}$};
			\node[latent, above=8mm of y] (kap) {$\kappa$};
			
			% arrows
			\draw[arr] (phi) -- (eta);
			\draw[arr] (eta) -- (f);
			\draw[arr] (f) -- (p);
			\draw[arr] (eps) -- (p);
			\draw[arr] (p) -- (y);
			\draw[arr] (kap) -- (y);
			
			% plate
			\node[plate, fit=(f)(p)(y), label=below right:{$d=1,\ldots,D$}] {};
			
		\end{tikzpicture}
		\caption{Directed acyclic graph showing the conditional dependence structure of our hierarchical NNGP model. Shaded nodes denote observed data and dashed nodes denote deterministic quantities.}
		\label{fig:dag_model}
	\end{figure}

	The likelihood for all studies is
	\begin{align*}
		p(y_{1:D}\mid f_{1:J, 1:D}, \epsilon_{1:J}, \kappa) = \prod_{d = 1}^{D}p(y_{d} \mid f_{1:J, d}, \epsilon_{1:J}, \kappa).
	\end{align*}
	The posterior of the model is
	\begin{align}
		\pi(f_{1:J, 1:D}, &\eta_{1:J}, \phi, \epsilon_{1:J}, \kappa\mid y_{1:D}) \nonumber \\ &\proptos p(y_{1:D}\mid f_{1:J, 1:D}, \epsilon_{1:J}, \kappa)p(f_{1:J, 1:D}\mid\eta_{1:J})p(\eta_{1:J}\mid\phi)p(\phi)p(\epsilon_{1:J}, \kappa), \label{eqn:posterior}
	\end{align}
	which naturally gives parameter blocks 
	\begin{align*}
		p(f_{1:J, 1:D}\mid y_{1:D}, \eta_{1:J}, \epsilon_{1:J}, \kappa) &\proptos p(y_{1:D}\mid f_{1:J, 1:D}, \epsilon_{1:J}, \kappa)p(f_{1:J, 1:D}\mid\eta_{1:J}) \\
		p(\eta_{1:J}\mid f_{1:J, 1:D}, \phi) &\proptos p(f_{1:J, 1:D}\mid\eta_{1:J})p(\eta_{1:J}\mid\phi) = \prod_{j = 1}^{J}p(f_{j, 1:D}\mid\eta_j)p(\eta_j\mid\phi)\\
		p(\phi\mid \eta_{1:J}) &\proptos p(\eta_{1:J}\mid\phi)p(\phi)\\
		p(\epsilon_{1:J}, \kappa\mid y_{1:D}, f_{1:J, 1:D}) &\proptos p(y_{1:D}\mid f_{1:J, 1:D}, \epsilon_{1:J}, \kappa)p(\epsilon_{1:J}, \kappa).
	\end{align*}
    Specifically, the blocks are the latent values $f_{1:J, 1:D}$, the parameters for the zero-mean NNGPs $\eta_{1:J}$, the hierarchical parameters $\phi$ and the likelihood parameters $\epsilon_{1:J}$ and $\kappa$. See Figure \ref{fig:dag_model} for a diagram showing the conditional dependence structure of the model. Note that each vector $\eta_j$ can be updated independently (and in parallel) conditional on $f_{j, 1:D}$ and $\phi$, and it is possible to draw $\phi$ directly from its full conditional posterior through choice of conjugate prior. 

    \section{Methodology} \label{sec:methods}

    \subsection{Sequential Monte Carlo}

	We use a likelihood tempering sequential Monte Carlo (SMC) algorithm for parameter inference of the spatio-temporal NNGP model. Let $\theta = \{f_{1:J, 1:D}, \eta_{1:J}, \phi, \epsilon_{1:J}, \kappa\}$. SMC samples from a sequence of $M$ distributions using a series of reweight, resample and mutation steps \citep{delmoral2006}. For our model, the sequence of distributions is
	\begin{equation}
		\begin{aligned}
			\pi_{m}(\theta_{m}\mid y_{1:D}) &\proptos \gamma_{m}(\theta_{m}, y_{1:D}) \\
			&= p(y_{1:D}\mid \theta)^{g_m}p(\theta) \\
			&= p(y_{1:D}\mid f_{1:J, 1:D}, \epsilon_{1:J}, \kappa)^{g_m}p(f_{1:J, 1:D}\mid\eta_{1:J})p(\eta_{1:J}\mid\phi)p(\phi)p(\epsilon_{1:J}, \kappa), 
		\end{aligned}
		\label{eqn:smc_target}
	\end{equation}
	for $m = 1, \ldots, M$, where $g_m = 0$ gives the prior and $g_m = 1$ gives the posterior in \eqref{eqn:posterior}. At each iteration, samples or particles are reweighted using the ratio of the new distribution in the sequence over the current distribution of the particles,
	\begin{align}
		w_m
		&=
		\frac{
			\pi_m(\theta_{m-1}\mid y_{1:D})
		}{
			\pi_{m-1}(\theta_{m-1}\mid y_{1:D})
		} 
		= p(y_{1:D}\mid f_{1:H, 1:D}, \epsilon_{1:J}, \kappa)^{g_m-g_{m-1}}. \label{eqn:smc_weights}
	\end{align}
	Given $N_{\theta}$ particles, the normalised weights are
	\begin{align*}
		W_m^n = \frac{w_m^n}{\sum_{i=1}^{N_{\theta}}w_m^i},
	\end{align*}
	for $n = 1,\ldots,N_{\theta}$. The effective sample size (ESS) can be approximated as
	\begin{align*}
		\widehat{\textrm{ESS}} = N_{\theta} \frac{(\sum_{n=1}^{N_{\theta}}w_m^n)^2}{\sum_{n=1}^{N_{\theta}}(w_m^n)^2},
	\end{align*}
	and the tempering parameter $g_m$ can be adapted by aiming for a specific ESS \citep{delmoral2006}.
	
	Once the particles have been reweighted, a resampling step is used to remove low weight particles and duplicate high weight particles. Finally, the mutation step diversifies the particles without changing their underlying distribution. See Algorithm \ref{alg:smc2} for more details.
	
	To mutate the particles, we use a Metropolis-within-Gibbs approach where the Metropolis-adjusted Langevin algorithm \citep[MALA; ][]{roberts1998,neal2011} is used to update $\eta_{1:J}$, $\epsilon_{1:J}$ and $\kappa$. The population parameters $\phi$ can also be updated using MALA, or Gibbs sampling through choice of conjugate prior. The NNGP function values $f_{1:J, 1:D}$ are updated using particle Gibbs with ancestor sampling \citep[PGAS; ][]{lindsten2014}. Since PGAS is used in the mutation step, the SMC algorithm falls within the SMC$^2$ framework \citep{chopin2012,duan2014}.
	
	An advantage of SMC is that the particles at each iteration can be used to adapt the proposals used in the mutation step, e.g.\ the MALA stepsize. This is discussed in more detail in Section \ref{sec:example_impl}. 
	
	\begin{algorithm}[htp]
		\small
		\begin{adjustwidth}{\algorithmicindent}{}
			\textbf{Input: } data $y_{1:D}$, number of parameter particles $N_{\theta}$, number of mutation iterations $R$ \\
			\textbf{Output: } posterior samples $\theta^{1:N_{\theta}}$
		\end{adjustwidth}
		\vspace{0.5em}
		
		\begin{algorithmic}[1]
			\LineComment{Initialise}
			\State Set $m = 1$ and $g_1 = 0$
			\For{$n=1$ to $N_{\theta}$} 
				\LineCommentIndent{Draw from prior}
				\State Sample $\phi^{n} \sim p(\phi)$, $\eta_{1:J}^{n} \sim p(\eta_{1:J}\mid\phi^n)$, $f_{1:J, 1:D}^{n} \sim p(f_{1:J, 1:D}\mid\eta_{1:J})$ and $\epsilon_{1:J}^{n}, \kappa^{n} \sim p(\epsilon_{1:J}, \kappa)$
				\State Set $W_1^{n} = \frac{1}{N_{\theta}}$
			\EndFor
			\vspace{0.5em}
			
			\While{$g<1$}
				\State Set $m = m + 1$
				\State Adapt $g_m$ by aiming for an $\widehat{\textrm{ESS}}$ of $N_{\theta}\slash 2$
				\vspace{0.5em}
				
				\LineComment{Reweight the particles from $\pi_{m-1}(\cdot)$ to $\pi_{m}(\cdot)$}
				\State Calculate the weights $W_m^n \proptos p(y_{1:D}\mid f_{1:J, 1:D}^n, \epsilon_{1:J}^n, \kappa^n)^{g_m-g_{m-1}}$ for $n = 1,\ldots, N_{\theta}$
				\vspace{0.5em}
				
				\LineComment{Resample}
				\State Sample the ancestor indices $a^{n}\sim\mathrm{Cat}(W_{d-1}^{1:N_{\theta}})$ 
				\State Set $\phi^n = \phi^{a^n}$, $\eta_{1:J}^n = \eta_{1:J}^{a^n}$, $f_{1:J,1:D}^n = f_{1:J,1:D}^{a^n}$, $\epsilon_{1:J}^n = \epsilon_{1:J}^{a^n}$ and $\kappa^n = \kappa^{a^n}$ for $n = 1,\ldots, N_{\theta}$
				\vspace{0.5em}
				
				\LineComment{Adapt proposals}
				\State Adapt the MALA and PGAS proposals (see Section \ref{sec:example_impl})
				\vspace{0.5em}
				
				\LineComment{Mutate}
				\For{$r=1$ to $R$} 
					\For{$n=1$ to $N_{\theta}$}
						\State Draw $\phi^n \sim p(\cdot\mid \eta_{1:J}^{n})$ 
						\State Update $\eta_{1:J}^n\mid f_{1:J, 1:D}^n, \phi^n$ using MALA
						\State Sample $f_{1:J, 1:D}^n\mid y_{1:D}, \eta_{1:J}^n, \epsilon_{1:J}^n, \kappa^n$ using PGAS
						\State Update $\epsilon_{1:J}^n, \kappa^n\mid y_{1:D}, f_{1:J, 1:D}^n$ using MALA
					\EndFor
				\EndFor
			\EndWhile
		\end{algorithmic}
		\caption{The SMC$^2$ Algorithm.}
		\label{alg:smc2}
	\end{algorithm}

	\subsection{Particle Gibbs}
	
	For ease of exposition, we omit dependence on $\epsilon_{1:J}$, $\kappa$ and $\eta_{1:J}$ in this section to emphasise inference of $f_{1:J, 1:d}$. Particle Gibbs \citep[PG; ][]{andrieu2010} is an SMC algorithm where one of the trajectories (the reference trajectory) is fixed throughout the iterations. As with SMC, a series of reweight, resample and mutation steps are applied to sample from a sequence of distributions, 
	\begin{align*}
		p_m(f_{1:J, 1:d}\mid y_{1:d}) &\proptos \gamma_m(f_{1:J, 1:d}, y_{1:d}) \\
		&= \gamma_m(f_{1:J, 1}, y_{1})\prod_{i=2}^{d}\gamma_m(f_{1:J, i}, y_{i}\mid f_{1:J, K(x_i)}) \\
		&= p(f_{1:J, 1})p(y_{1}\mid f_{1:J, 1})^{g_m}\prod_{i=2}^{d}p(f_{1:J,i}\mid f_{1:J, K(x_i)})p(y_{i}\mid f_{1:J, i})^{g_m},
	\end{align*}
	for $d = 1, \ldots, D$. Particle Gibbs also falls into the pseudo-marginal framework in that it is exact Gibbs sampling on an extended space \citep{andrieu2009} and targets the correct distribution marginally. We follow \citet{finke2023} and \citet{corenflos2024} and introduce additional auxiliary variables or pseudo-observations $u_{1:D}$ to guide proposals for $f_{1:J, 1:d}$. These variables are marginalised out in the weights.
	
	Let $N_{\textrm{PG}}$ denote the number of particles used in PG, $a_{d-1}^{1:N_{\textrm{PG}}}$ the ancestor indices and $k_d\in\{1,\dots,N_{\textrm{PG}}\}$ the index of the reference trajectory. Let $i\ne k_d$ and superscript $-k_d$ mean all particle indices except $k_d$, e.g.\ $f_{1:J, d}^{-k_d} = (f_{1:J, d}^{1}, \ldots, f_{1:J, d}^{k_d-1}, f_{1:J, d}^{k_d+1}, \ldots, f_{1:J, d}^{N_{\textrm{PG}}})$. The proposal for $f_{1:J, d}^{-k_d}$ and $u_d$ is \citep{corenflos2024} 
	\begin{align*}
		q_d\left(f_{1:J, d}^{-k_d},u_d\middle| f_{1:J, d}^{k_d},f_{1:J, K(x_d)}^{1:N_{\textrm{PG}}}\right)
		=
		q_d\left(u_d\middle| f_{1:J, d}^{k_d}, f_{1:J, K(x_d)}^{k_d}\right)
		\prod_{i\neq k_d} q_d\left(f_{1:J, d}^i \middle| f_{1:J, K(x_d)}^i,u_d\right),
	\end{align*}
	where
	\begin{align}
		q_d\left(u_d\middle| f_{1:J, d}^{k_d}, f_{1:J, K(x_d)}^{k_d}\right) = \mathcal{N}\left(u_d\middle| f_{1:J, d}^{k_d} + \frac{\epsilon^2}{2}S_d\phi_d(f_{1:J, K(x_d)}^{k_d}), \frac{\epsilon^2}{2}S_d\right). \label{eqn:u_prop}
	\end{align}
	The marginal proposal for $f_{1:J, d}^{-k_d}$ is then
	\begin{align}
		q_d(f_{1:J, d}^{-k_d}\mid f_{1:J, d}^{k_d}, f_{1:J, K(x_d)}^{1:N_{\textrm{PG}}})= \int{q_d(f_{1:J, d}^{-k_d},u_d\mid f_{1:J, d}^{k_d}, f_{1:J, K(x_d)}^{1:N_{\textrm{PG}}})}du_d. \label{eqn:marg_f_prop}
	\end{align}
	If the transition kernel $q_d\left(f_{1:J, d} \middle| f_{1:J, K(x_d)},u_d\right)$ is Gaussian with a covariance matrix that does not depend on $f_{1:J, K(x_d)}$, then the integral in \eqref{eqn:marg_f_prop} is analytically tractable.

	At each iteration, the non-reference particles $f_{1:J, d}^{-k_d}$ and reference index $k_{d}$ are sampled as part of a Gibbs update on the extended target 
	\begin{align*}
		\tilde\pi_d(k_d,f_{1:J, d}^{-k_d}\mid f_{1:J, d}, f_{1:J, K(x_d)}^{1:N_{\textrm{PG}}}) &\proptos
		\gamma_m(f_{1:J, d}^{k_d}, y_{d}\mid f_{1:J, K(x_d)}^{k_d})
		q_d(f_{1:J, d}^{-k_d}\mid f_{1:J, d}^{k_d},f_{1:J, K(x_d)}^{1:N_{\textrm{PG}}})
	\end{align*}
	where
	\begin{align*}
		\tilde\pi_d(k_d\mid f_{1:J, d}^{1:N_{\textrm{PG}}}, f_{1:J, K(x_d)}^{1:N_{\textrm{PG}}}) &\proptos
		\gamma_m(f_{1:J, d}^{k_d}, y_{d}\mid f_{1:J, K(x_d)}^{k_d})
		q_d(f_{1:J, d}^{-k_d}\mid f_{1:J, d}^{k_d},f_{1:J, K(x_d)}^{1:N_{\textrm{PG}}})
	\end{align*}
	and
	\begin{align*}
		\tilde\pi_d(f_{1:J, d}^{-k_d}\mid k_d, f_{1:J, d}^{k_d}, f_{1:J, K(x_d)}^{1:N_{\textrm{PG}}}) &\proptos
		q_d(f_{1:J, d}^{-k_d}\mid f_{1:J, d}^{k_d},f_{1:J, K(x_d)}^{1:N_{\textrm{PG}}}).
	\end{align*}
	See Algorithm \ref{alg:pg} for more details.

	\begin{algorithm}[htp]
		\small
		\begin{adjustwidth}{\algorithmicindent}{}
			\textbf{Input: } reference trajectory $f_{1:J, 1:D}$, number of particles $N_{\textrm{PG}}$ \\
			\textbf{Output: } selected trajectory $f_{1:J, 1:D}^{k_D}$ 
		\end{adjustwidth}
		\vspace{0.5em}
		
		\begin{algorithmic}[1]
			\For{$d=1$ to $D$} 
			
			\State Sample $k_d \sim \mathrm{Unif}\{1,\dots,N_{\textrm{PG}}\}$ and set $f_{1:J, d}^{k_d}=f_{1:J, d}$
			\vspace{0.5em}
			
			\LineComment{Resample and update ancestry}
			\If{$d>1$}
			\State Set $a_{d-1}^{k_d}=k_{d-1}$ \label{alg_step:ref_index}
			\State Sample ancestor indices $a_{d-1}^n \sim \mathrm{Cat}(W_{d-1}^{1:N_{\textrm{PG}}})$ for $n\neq k_d$
			\State Set $f_{1:J, 1:d-1}^n = f_{1:J, 1:d-1}^{a_{d-1}^n}$
			\EndIf
			\vspace{0.5em}
			
			\LineComment{Mutate}
			\State Sample $f_{1:J, d}^{-k_d}$ from $q_d(f_{1:J, d}^{-k_d}\mid f_{1:J, d}^{k_d},f_{1:J, K(x_d)}^{1:N_{\textrm{PG}}})$:
			\begin{itemize}
				\item Draw $u_d \sim q_d(\cdot\mid f_{1:J, d}^{k_d}, f_{1:J, K(x_d)}^{k_d})$
				\item Draw $f_{1:J, d}^{-k_d} \sim \prod_{n\neq k_d} q_d\left(f_{1:J, d}^{n} \mid f_{1:J, K(x_d)}^{n},u_d\right)$ and discard $u_d$
			\end{itemize}
			\vspace{0.5em}
			
			\LineComment{Reweight}
			\State Compute weights $W_d^n \proptos \gamma(f_{1:J, d}^{n}, y_{d}\mid f_{1:J, K(x_d)}^{n})q_d(f_{1:J, d}^{-n}\mid f_{1:J, d}^{n},f_{1:J, K(x_d)}^{1:N_{\textrm{PG}}})$ for $n=1,\ldots,N_{\textrm{PG}}$
			\EndFor
			\State Draw $k_D \sim \mathrm{Cat}(W_{D}^{1:N_{\textrm{PG}}})$ %and set $k_{d-1} = a_{d-1}^{k_d}$ for $d = D, \ldots, 2$
		\end{algorithmic}
		\caption{The Particle Gibbs Algorithm.}
		\label{alg:pg}
	\end{algorithm}

	\subsection{PG Proposal Distributions}
	
	We now discuss our three choices of $q_d\left(u_d\middle| f_{1:J, d}^{k_d}, f_{1:J, K(x_d)}^{k_d}\right)$ and $q_d\left(f_{1:J, d} \mid f_{1:J, K(x_d)},u_d\right)$.
	
	\subsubsection{Bootstrap}
	The first choice of $q_d$ sets
	\begin{align*}
		q_d\left(f_{1:J, d} \middle| f_{1:J, K(x_d)}, u_d\right) = p\left(f_{1:J,d}\middle| f_{1:J, K(x_d)}\right),
	\end{align*}
	giving
	\begin{align*}
		q_d(f_{1:J, d}^{-k_d}\mid f_{1:J, d}^{k_d},f_{1:J, K(x_d)}^{1:N_{\textrm{PG}}})
		&= \int{q(u_d\mid f_{1:J, d}^{k_d}, f_{1:J, K(x_d)}^{k_d})\prod_{i\neq k_d} p(f_{1:J,d}^{i}\mid f_{1:J, K(x_d)}^{i})}du_d \\
		&= \prod_{i\neq k_d} p(f_{1:J,d}^{i}\mid f_{1:J, K(x_d)}^{i}),
	\end{align*}
	and
	\begin{align*}
		W_d^n &\proptos \gamma_m(f_{1:J, d}^{n}, y_{d}\mid f_{1:J, K(x_d)}^{n})
		q_d(f_{1:J, d}^{-n}\mid f_{1:J, d}^{n},f_{1:J, K(x_d)}^{1:N_{\textrm{PG}}}) \\
		&= p(f_{1:J,d}^{n}\mid f_{1:J, K(x_d)}^{n})p(y_{d}\mid f_{1:J, d}^{n})^{g_m}\prod_{i\neq n} p(f_{1:J,d}^{i}\mid f_{1:J, K(x_d)}^{i}) \\
		&\proptos p(y_{d}\mid f_{1:J, d}^{n})^{g_m},
	\end{align*}
	which corresponds to the standard bootstrap proposal.

	\subsubsection{Random Walk}
	The second choice sets $\phi_d(\cdot) = 0$ in \eqref{eqn:u_prop} giving 
	\begin{align*}
		q_d\left(u_d\middle| f_{1:J, d}^{k_d}, f_{1:J, K(x_d)}^{k_d}\right) = \mathcal{N}\left(u_d \middle| f_{1:J, d}^{k_d}, \frac{\epsilon^2}{2}S_d\right),
	\end{align*}
	and sets
	\begin{align*}
		q_d\left(f_{1:J, d} \middle| f_{1:J, K(x_d)},u_d\right) = \mathcal{N}\left(f_{1:J, d} \middle| u_d, \frac{\epsilon^2}{2}S_d\right).
	\end{align*}
	This falls under the Particle-MALA algorithm of \citet{corenflos2024} with the gradient information switched off. The marginal proposal is
	\begin{align*}
		q_d(f_{1:J, d}^{-k_d}\mid f_{1:J, d}^{k_d},f_{1:J, K(x_d)}^{1:N_{\textrm{PG}}})
		&= \int{\mathcal{N}\left(u_d \middle| f_{1:J, d}^{k_d}, \frac{\epsilon^2}{2}S_d\right)\prod_{i\neq k_d} \mathcal{N}\left(f_{1:J, d}^{i} \middle| u_d, \frac{\epsilon^2}{2}S_d\right)}du_d \\
		&= \int{\prod_{i=1}^N \mathcal{N}\left(u_d \middle| f_{1:J, d}^{i}, \frac{\epsilon^2}{2}S_d\right)}du_d,
	\end{align*}
	which gives incremental weights
	\begin{align*}
		W_d^n &\proptos \gamma_m(f_{1:J, d}^{n}, y_{d}\mid f_{1:J, K(x_d)}^{n})
		q_d(f_{1:J, d}^{-n}\mid f_{1:J, d}^{n},f_{1:J, K(x_d)}^{1:N_{\textrm{PG}}}) \\
		&= \gamma_m(f_{1:J, d}^{n}, y_{d}\mid f_{1:J, K(x_d)}^{n})
		\int{\prod_{i=1}^N \mathcal{N}\left(u_d \middle| f_{1:J, d}^{i}, \frac{\epsilon^2}{2}S_d\right)}du_d \\
		&\proptos\gamma_m(f_{1:J, d}^{n}, y_{d}\mid f_{1:J, K(x_d)}^{n}).
	\end{align*}
	This method reduces to the Particle-RWM method of \citet{finke2023}.

	\subsubsection{MALA}
	The final choice of $q_d$ is the Particle-mGRAD proposal of \citet{corenflos2024}, which sets
	\begin{align*}
		\phi_d(f_{1:J, K(x_d)}) = \nabla\log{p(y_{d}\mid f_{1:J, d})^{g_m}} = g_m\nabla\log{p(y_{d}\mid f_{1:J, d})}
	\end{align*}
	and
	\begin{align*}
		q_d\left(f_{1:J, d} \middle| f_{1:J, K(x_d)}, u_d\right) 
		&\proptos q_d\left(f_{1:J, d} \middle| f_{1:J, K(x_d)}\right)\mathcal{N}\left(f_{1:J, d} \middle| u_d, \frac{\epsilon^2}{2}S_d\right) \\
		&= \mathcal{N}\left(f_{1:J,d}\mid m_d(f_{1:J,K(x_d)}),C_d\right)\mathcal{N}\left(f_{1:J, d} \middle| u_d, \frac{\epsilon^2}{2}S_d\right) \\
		&= \mathcal{N}\left(f_{1:J,d} \mid v_d + H_d u_d, D_d \right),
	\end{align*}
	where $q_d\left(f_{1:J, d} \middle| f_{1:J, K(x_d)}\right)$ is defined in \eqref{eqn:transition} and
	\begin{align*}
		H_d = C_d\left(C_d+\frac{\epsilon^2}{2}S_d\right)^{-1}, \quad
		v_d = (I-H_d)\,m_d(f_{1:J,K(x_d)}), \quad
		D_d = \frac{\epsilon^2}{2}S_d\left(C_d+\frac{\epsilon^2}{2}S_d\right)^{-1}C_d.
	\end{align*}
	Let $\phi_d^{k_d} = \phi_d\left(f_{1:J, K(x_d)}^{k_d}\right)$. The proposal is 
	\begin{align*}
		q_d(f_{1:J, d}^{-k_d}\mid f_{1:J, d}^{k_d},f_{1:J, K(x_d)}^{1:N_{\textrm{PG}}})
		\proptos H_{d, \phi_d^{k_d}}\left(f_{1:J, d}^{k_d}, v_d^{k_d}, \bar{f}_{1:J, d}, \bar{v}_d\right) \\
		\bar{f}_{1:J, d} = \frac{1}{N_{\textrm{PG}}}\sum_{n=1}^{N_{\textrm{PG}}}f_{1:J, d}^{n}, \quad
		\bar{v}_{d} = \frac{1}{N_{\textrm{PG}}}\sum_{n=1}^{N_{\textrm{PG}}}v_{d}^{n},
	\end{align*}
	where
	\begin{align*}
		\log H_{d,\phi_d^{k_d}}&\left(f_{1:J,d}^{k_d},v_d^{k_d},\bar f_{1:J,d},\bar v_d\right) \\
		&= \frac{1}{2}\left(f_{1:J,d}^{k_d} - v_d^{k_d}\right)^{\top}\left(D_d^{-1} + G_d\right)\left(f_{1:J,d}^{k_d} - v_d^{k_d}\right) \\
		&\quad -\frac{1}{2} (N_{\textrm{PG}}-1)\left(f_{1:J,d}^{k_d} + \phi_d^{k_d}\right)^{\top}H_d^{\top}\left(D_d^{-1} - (N_{\textrm{PG}}-1)G_d\right)H_d\left(f_{1:J,d}^{k_d} + \phi_d^{k_d}\right) \\
		&\quad -N_{\textrm{PG}}(\bar{f}_{1:J,d} - \bar{v}_d)^{\top}G_d\left[\left(f_{1:J,d}^{k_d} - v_d^{k_d}\right) - \left(D_d^{-1} - (N_{\textrm{PG}}-1)G_d\right)H_d\left(f_{1:J,d}^{k_d} + \phi_d^{k_d}\right)\right] \\
		&\quad -\left(f_{1:J,d}^{k_d} - v_d^{k_d}\right)^{\top}\left(D_d^{-1} - (N_{\textrm{PG}}-1)G_d\right)H_d\left(f_{1:J,d}^{k_d} + \phi_d^{k_d}\right),
	\end{align*}
	and 
	\begin{align*}
	G_d = \left(D_d + (N_{\textrm{PG}}-1)H_d \frac{\epsilon^2}{2}S_d H_d^{\top} \right)^{-1}H_d \frac{\epsilon^2}{2}S_d H_d^{\top}D_d^{-1}.
	\end{align*}
	The incremental weights are
	\begin{align*}
		W_d^n &\proptos \gamma(f_{1:J, d}^{n}, y_{d}\mid f_{1:J, K(x_d)}^{n})q_d(f_{1:J, d}^{-n}\mid f_{1:J, d}^{n},f_{1:H, K(x_d)}^{1:N_{\textrm{PG}}}) \\
		&\proptos \gamma(f_{1:J, d}^{n}, y_{d}\mid f_{1:J, K(x_d)}^{n})H_{d, \phi_d^{n}}\left(f_{1:J, d}^{n}, v_d^{n}, \bar{f}_{1:J, d}, \bar{v}_d\right).
	\end{align*}
	See \citet{corenflos2024} for more details. 
	
	\subsubsection{Complexity} \label{sssec:cost}
	
	For a fixed number of particles, the cost of particle Gibbs is linear in the number of observations $D$ \citep{corenflos2024}. The cost of the bootstrap and random walk proposals is linear in $J$ as it only requires simulating\slash evaluating the $J$ NNGPs once per iteration. In general, the cost of the MALA proposal \citep[Particle-mGRAD in][]{corenflos2024} is cubic in the state dimension $J$, but since the NNGP covariance matrix $C_d$ is diagonal and does not depend on $f_{1:J, K(x_d)}$, the cost of this proposal also reduces to linear in $J$.

	\subsubsection{Adaptation}
	
	We use a mixture of the bootstrap, random walk and MALA proposals in particle Gibbs. The bootstrap proposal works well when the prior is close to the target, i.e.\ $\gamma_m(f_{1:J, d}, y_{d}\mid f_{1:J, K(x_d)}) \approx p(f_{1:J,d}\mid f_{1:J, K(x_d)})$, which may occur if $f_{1:J, d}$ is close to its neighbors $f_{1:J, K(x_d)}$ or if $g_m \approx 0$. Conversely, the MALA proposal has good performance if the likelihood is very informative relative to the prior. Since $g_1 = 0$ in the SMC algorithm, the bootstrap proposal is initially used for all points $d = 1, \ldots, D$.
	
	The main motivation for the three proposals is that the bootstrap proposal is computationally cheaper than the random walk proposal, and the random walk proposal is computationally cheaper than the MALA proposal. Additionally, the MALA proposal behaves more like the bootstrap proposal when the prior is informative relative to the likelihood \citep{corenflos2024}. In this setting, it is both unclear how to adapt the stepsize parameter $\epsilon$, as we find it has little impact on the acceptance rate, and it is computationally cheaper to use the bootstrap proposal directly. Section \ref{sec:example_impl} describes how we adapt the stepsize automatically within the SMC$^2$ algorithm.
	
	To test how informative the likelihood is relative to the prior, we compare the squared Euclidean norms of the gradient of the log-likelihood and the gradient of the log-prior, and take the average over the $N_{\theta}$ particles, 
	\begin{align}
		s_d &= \frac{1}{N_{\theta}}\sum_{n=1}^{N_{\theta}}\frac{\|\nabla\log{p(y_{i}\mid f_{1:J, i}^n)^{g_m}}\|_2^2}{\|\nabla\log{\mathcal{N}\left(f_{1:J,d}^n\mid m_d(f_{1:J,K(x_d)}^n),C_d\right)}\|_2^2} \nonumber \\
		&= \frac{1}{N_{\theta}}\sum_{n=1}^{N_{\theta}}\frac{\|\phi_d^n\|_2^2}{\left(f_{1:J,d}^n - m_d(f_{1:J,K(x_d)}^n)\right)^{\top}C_d^{-2}\left(f_{1:J,d}^n - m_d(f_{1:J,K(x_d)}^n)\right)}. \label{eqn:mala_proposal_metric}
	\end{align}
	The following procedure is used to determine the proposals at each iteration. %

	\paragraph{Bootstrap to Random Walk}
	
	\begin{enumerate}
		\item Run PG for each particle and calculate the average movement for each point $d$, without mutating the particles. The average movement is defined as 
		\begin{align}
			\alpha_{d, \textrm{PG}} = \frac{1}{N_{\theta}}\sum_{n=1}^{N_{\theta}}I(f_{1:J, d}^n \ne \tilde{f}_{1:J, d}^n), \label{eqn:pgas_ar}
		\end{align}
		where $f_{1:J, d}$ is the current value and $\tilde{f}_{1:J, d}$ is the new value drawn using PG. In this step, the new draws $\tilde{f}_{1:J, d}^{1:N_{\theta}}$ are discarded.
		\item If $\alpha_{d, \textrm{PG}}$ is below some threshold and the current proposal for $d$ is the bootstrap, then the proposal is switched to random walk.
	\end{enumerate}
	
	\paragraph{Random Walk to MALA}
	
	\begin{enumerate}
		\item[3.] Calculate $s_d$ as in \eqref{eqn:mala_proposal_metric} for all $d$ that currently use a random walk proposal. If $s_d > 1$, then switch the proposal for $d$ to MALA.
	\end{enumerate}

	\subsection{Ancestral Sampling}
	
	While the PG algorithm targets $p(f_{1:J, 1:d}\mid y_{1:d})$ for any choice of $N_{\textrm{PG}}$ \citep{andrieu2010}, the mixing is generally poor due to path degeneracy \citep{lindsten2013,chopin2015}, i.e.\ the collapse of particle genealogies caused by repeated resampling, which leads to most or all of the particles sharing a common ancestor at some point.
	
	For Markov models, path degeneracy can be mitigated through the use of backward sampling \citep{whiteley2010}. In our setting, particle Gibbs with ancestor sampling \citep[PGAS; ][]{lindsten2014} can be used instead. PGAS samples a new index for the reference path at each iteration with weights 
	\begin{align*}
		\tilde{w}^n_{d-1|D} &= w_{d-1}^n\frac{\gamma(f_{1:J, 1:d-1}^n, f_{1:J, d:D}, y_{1:d})}{\gamma(f_{1:J, 1:d-1}^n, y_{1:d-1})} = \gamma(f_{1:J, d:D}, y_{1:d}\mid f_{1:J, 1:d-1}^n)  \\
		&= w_{d-1}^n  \prod_{i=d}^{D}p(f_{1:J,i}\mid f_{1:J, K(x_i)}^n)p(y_{i}\mid f_{1:J, i})^{g_m} \\
		&\proptos w_{d-1}^n  \prod_{i=d}^{D}p(f_{1:J,i}\mid f_{1:J, K(x_i)}^n). 
	\end{align*}
	Here, $w_{d-1}^n$ may be seen as the prior probability of $f_{1:J, 1:d-1}^n$ and $\gamma(f_{1:J, d:D}, y_{1:d}\mid f_{1:J, 1:d-1}^n)$ as the likelihood of $f_{1:J, d:D}$ descending from $f_{1:J, 1:d-1}^n$. The weights can equivalently be written as
	\begin{align}
		\tilde{w}^n_{d-1|D} &= w_{d-1}^n\gamma(f_{1:J, F(x_{d:D})}, y_{1:d}\mid f_{1:J, 1:d-1}^n) \nonumber \\
		&\proptos w_{d-1}^n  \prod_{i \in F(x_{d:D})}p(f_{1:J,i}\mid f_{1:J, K(x_i)}^n),\label{eqn:as_weights}
	\end{align}
	where $F(x_{d:D})$ is the set of indices in $\{d,\ldots,D\}$ that have neighbors in $\{1,\ldots,d-1\}$.
	
	While $p(f_{1:J,d}\mid f_{1:J, K(x_d)})$ has a constant cost, the computational complexity of \eqref{eqn:as_weights} is $O(D)$, which gives a quadratic complexity in $D$ for PGAS. To reduce the cost of \eqref{eqn:as_weights}, \citet{lindsten2014} propose a lag-based approximation, which truncates the product in \eqref{eqn:as_weights} to a fixed number of factors $\ell$.	In the context of NNGPs, this is equivalent to taking the product over the first $\ell$ elements of $F(x_{d:D})$, denoted $F_{1:\ell}(x_{d:D})$,
	\begin{align}
		\tilde{w}_{d-1\mid D, \ell}^n &\proptos w_{d-1}^n \prod_{i \in F_{1:\ell}(x_{d:D})} p\left(f_{1:J, i}\mid f_{1:J, K(x_i)}^n\right), \quad \tilde{W}_{d-1\mid D, \ell}^n = \frac{\tilde{w}_{d-1\mid D, \ell}^n}{\sum_{i = 1}^{N_{\textrm{PG}}}\tilde{w}_{d-1\mid D, \ell}^i}. \label{eqn:pgas_weights_ref_index}
	\end{align}
	
	We use the adaptive approach of \citet{lindsten2014} to set $\ell$ at each iteration of PGAS. Specifically, they iteratively calculate the total variation distance between $\tilde{W}_{d-1\mid D, \ell}^n$ and $\tilde{W}_{d-1\mid D, \ell-1}^n$, for $\ell = 0, \ldots, |F(x_{d:D})|$, where $|F(x_{d:D})| \le D-d+1$,
	\begin{align*}
		\epsilon_{\ell} = \frac{1}{2}\sum_{i=1}^{N_{\textrm{PG}}}\left|\tilde{W}_{d-1\mid D, \ell}^n - \tilde{W}_{d-1\mid D, \ell-1}^n\right|,
	\end{align*}
	and use the exponential moving average as a stopping rule
	\begin{align*}
		\bar{\epsilon}_{\ell} = \nu \bar{\epsilon}_{\ell-1} + (1-\nu)\epsilon_{\ell} < \tau, \quad \nu,\tau\in [0,1].
	\end{align*}
	\citet{lindsten2014} find that the default values of $\nu = 0.1$ and $\tau = 0.01$ give good results for a range of examples, and we use these values for all of our examples in Section \ref{sec:example}. To implement PGAS, step \ref{alg_step:ref_index} of Algorithm \ref{alg:pg} is replaced with 
	\begin{enumerate}
		\item Calculate $\tilde{W}_{d-1\mid D, \ell}^n$ as in \eqref{eqn:pgas_weights_ref_index} for $n = 1,\ldots,N_{\textrm{PG}}$ using the adaptive approach of \citet{lindsten2014}.
		\item Sample $a_{d-1}^{k_d} \sim \Cat(\tilde{W}_{d-1\mid D, \ell}^{1:N_{\textrm{PG}}})$ for $n = 1,\ldots,N_{\textrm{PG}}$.
	\end{enumerate}

	In practice, we find that the cost of calculating the weights in \eqref{eqn:pgas_weights_ref_index} using the adaptive approach is less than $O(k)$.

	\section{Case Study: Antimalarial Drug Resistance} \label{sec:example}
	
	Antimalarial drug resistance is an ongoing problem in Sub-Saharan Africa \citep{who2025}. While difficult to quantify directly, resistance to particular antimalarial drugs has been linked with specific mutations in the parasite genome \citep{who2025a}. Combinations of mutations (or haplotypes) are associated with a greater degree of resistance compared to single marker mutations \citep{kublin2002}. 
	
	Despite widespread resistance to the antimalarial drug sulfadoxine-pyrimethamine (SP) as a malaria clearing agent, its use during pregnancy has been associated with higher birth weights and lower infant mortality compared to no intervention \citep{eisele2012,vaneijk2019}. As a result, the World Health Organization (WHO) currently recommends SP as an intermittent preventive treatment during pregnancy (IPTp-SP) \citep{who2022}. Resistance to SP is associated with mutations in the \textit{dhps} and \textit{dhfr} genes, with the quintuple mutation \textit{dhfr}+\textit{dhps} 51I-59R-108N-437G-540E associated with a high level of resistance to SP as a clearing agent. More recently, reduced effectiveness of IPTp-SP in improving birth outcomes has been linked to the sextuple mutation \textit{dhfr}+\textit{dhps} 51I-59R-108N-437G-540E-581G \citep{vaneijk2019}. 
	
	Due to computational limitations, focus has been on inferring the prevalence of the single marker mutation \textit{dhps} 581G \citep{flegg2013,flegg2022} and more recently, the triple mutation \textit{dhps} 437G-540E-581G \citep{foo2024}. In this section, we show that our model gives similar results to \citet{foo2024} when applied to a 3 \textit{dhps} marker dataset, and that it easily scales to 6 markers, allowing the prevalence of the sextuple mutation to be inferred directly for the first time.

	\subsection{Model and Data}
	
	We use molecular surveillance data that were collated by the Worldwide Antimalarial Resistance Network into their Molecular Surveyor dataset, a living systematic review of the literature \citep{iddo2026}. This is the same dataset used in \cite{foo2024}, and includes studies conducted from 2000 to 2020. The dataset has prevalence data for markers \textit{dhps} A437G, K540E and A581G and \textit{dhfr} N51I, C59R and S108N in Sub-Saharan Africa. We filter the dataset to include only studies where the timing of surveillance was a window of less than 3 years. We count mixed infections (where both wild-type and mutant alleles were present in a single infection) as mutations and remove any studies with obvious inconsistencies in counts, e.g.\ where multimarker counts are inconsistent with single marker counts. After processing, the data include studies from 2000 to 2017.
	
	For each study $d = 1,\ldots,D$, we have covariates $x_d = (u_d, v_d, t_d, r_d)$, where $u_d$ is the latitude, $v_d$ is the longitude, $t_d$ is the median year of the study and $r_d$ is the \textit{Plasmodium falciparum} parasite rate \citep[retrieved from the][]{map2025}.

    We follow \citet{foo2024} and assume the mean is a linear function of the parasite rate $r_d$, and use a sphere-Gneiting covariance function \citep{porcu2021} as the kernel 
	\begin{align*}
		m_j(x_d) &= \mu_j + \beta_j r_d \\
		k_j(x_d, x_d') &= s_j^2	\left(1 + \frac{(t_d - t_{d'})^2}{\tau_j^2}	+ \frac{d_{\textrm{GC}}(x_d, x_d')}{\delta_j}\right)^{-1} + \sigma_j^2 I(d = d').
	\end{align*}
	Unlike \citet{foo2024}, however, we allow the noise variance $\sigma_j^2$ to vary between NNGPs. Here $d_{\textrm{GC}}(a, b)$ is the Great Circle or Haversine distance between $a$ and $b$ in degrees, and the unknown parameters are $\eta_j = (\alpha_j, \nu_j, \tau_j, \delta_j)$ and $\epsilon_j = (\mu_j, \beta_j)$. The spatial and temporal distances are divided by the median non-zero distance between all points. For $d_{\textrm{GC}}(a, b)$, the median is $24$ degrees and for $(t_d - t_{d'})$, the median is $4$ years. Neighbors are chosen to minimise the sum of the two scaled distances, $\nicefrac{(t_d - t_{d'})^2}{4^2} + \nicefrac{d_{\textrm{GC}}(x_d, x_d')}{24}$. In all examples, $k = 20$ neighbors are used. To improve parameter identifiability, we enforce strong hierarchical pooling on $\tau_{1:J}$ and $\delta_{1:J}$. The priors are
	\begin{align*}
		p(\epsilon_{1:J}) =& \prod_{j=1}^{J}\operatorname{Normal}(\mu_j\mid 0, 2^2)\operatorname{Normal}(\beta_j\mid 0, 2^2), \\
		p(\eta_{1:J}\mid\phi) =& \prod_{j=1}^{J} \operatorname{InverseGamma}(s_j\mid 5, \beta_{s}) \operatorname{InverseGamma}(\sigma_j \mid 5, \beta_{\sigma}^2) \\
		&\quad \times\operatorname{InverseGamma}(\tau_j\mid 50, \beta_{\tau})\operatorname{InverseGamma}(\delta_j\mid 50, \beta_{\delta})
	\end{align*}
	where $\phi = (\beta_{s}, \beta_{\sigma}, \beta_{\tau}, \beta_{\delta})$ and
	\begin{align*}
		p(\phi) &= \operatorname{Gamma}(\beta_{s}\mid 3, 2) \operatorname{Gamma}(\beta_{\sigma}\mid 1, 1.5) \operatorname{Gamma}(\beta_{\tau}\mid 2, 20) \operatorname{Gamma}(\beta_{\delta}\mid 2, 20).
	\end{align*}
	As the priors for $\phi$ are conjugate, exact Gibbs sampling is used to update $\phi$.
	
	We use the Beta-Binomial likelihood in \eqref{eqn:likelihood}, with composite weights chosen to reduce redundancy. Recall that the configuration matrix $A_d\in\{0,1\}^{h_d, H}$ encodes which haplotypes a study has information on, such that $\tilde{p}_d = A_dp_d$. Let $B_d$ be the row-centered configuration matrix, $B = A(I_H - H^{-1}\boldsymbol{1}_{H}\boldsymbol{1}_{H}^{\top})$, where $I_H$ is the $H$ by $H$ identity matrix and $\boldsymbol{1}_{H}$ is a $H$-dimensional vector of ones. The likelihood weights are given by the leverage of $B_d$, 
	\begin{align*}
		w_{i,d} = b_{i,d}^{\top}(B_d^{\top}B_d)^{+}b_{i,d}, 
	\end{align*}
	where $b_{i,d}$ is the $i$th row of $B_d$ and $(B_d^{\top}B_d)^{+}$ is the Moore-Penrose pseudo-inverse of $B_d^{\top}B_d$. Each weight is between $0$ and $1$, weight is shared between redundant counts, and the sum of the weights is equal to the rank of the matrix $B_d$. As an example, the configuration matrix associated with observations \textit{dhps} 437G, A437, 540E, 437G-540E, 437G-540E-A581 and 437G-540E-581G,
    $$
	A_d = 
	\begin{pNiceMatrix}[first-row,first-col]
		& 000 & 100 & 010 & 110 & 001 & 101 & 011 & 111 \\
		(\phantom{-}1,-1,-1) & 0 & 1 & 0 & 1 &	0 & 1 &	0 & 1 \\
		(\phantom{-}0,-1,-1) & 1 & 0 & 1 & 0 &	1 & 0 &	1 & 0 \\
		(-1,\phantom{-}1,-1) & 0 & 0 & 1 & 1 &	0 & 0 &	1 & 1 \\
		(\phantom{-}1, \phantom{-}1, -1) & 0 & 0 & 0 & 1 &	0 & 0 &	0 & 1 \\
		(\phantom{-}1, \phantom{-}1, \phantom{-}0) & 0 & 0 & 0 & 1 & 0 & 0 & 0 & 0 \\
		(\phantom{-}1, \phantom{-}1, \phantom{-}1) & 0 & 0 & 0 & 0 & 0 & 0 & 0 & 1 
	\end{pNiceMatrix}.
	$$
	gives weights $w_d = (\frac{1}{2}, \frac{1}{2}, 1, \frac{2}{3}, \frac{2}{3}, \frac{2}{3})^{\top}$. So, the marginal likelihoods of \textit{dhps} 437G and A437 will be raised to a power of $1\slash 2$, the marginal likelihood of \textit{dhps} 540E will be raised to a power of $1$, and the marginal likelihoods of \textit{dhps} 437G-540E, 437G-540E-A581 and 437G-540E-581G will be raised to a power of $2\slash 3$.

	\subsection{Implementation} \label{sec:example_impl}
	
	We consider a dataset with both the 3 \textit{dhps} markers (giving $H=8$ haplotypes) and the 3 \textit{dhps} and 3 \textit{dhfr} markers (giving $H=64$ haplotypes). The number of observations in each dataset is $D = 234$ and $D = 243$ respectively. As a baseline, we also simulate synthetic data from the model for the same studies as in the 3 marker dataset. The code is implemented in Julia and is available at \url{https://github.com/imkebotha/particle-gibbs-for-NNGP-models}.
	
	SMC$^2$ is used for parameter inference. In all examples, we use $N_{\theta}=1000$ particles in SMC$^2$ and $N_{\textrm{PG}} = 20$ particles in PGAS. Systematic resampling is used for SMC$^2$ and PGAS \citep{douc2005}. The tempering parameters are adapted to achieve an approximate ESS of $500$, and the number of mutation steps is set based on the minimum acceptance rate across the parameter blocks from the previous iteration, i.e. $R = \lceil 10 \alpha^{-1}_{\textrm{min}}\rceil$, where $\alpha_{\textrm{min}}$ is the minimum overall acceptance rate (averaged across the $R$ iterations) from the previous mutation step. In practice, we find that the adaptation of the stepsizes generally leads to $\alpha_{\textrm{min}} \approx 0.5$ and thus $R \approx 20$ across all iterations.
	
	Preconditioned MALA is used to update $\eta_{1:J}$, $\epsilon_{1:J}$ and $\kappa$. Each $\eta_j$ is updated independently of the other $\eta_k$, $k\ne j$. The preconditioning matrix is the sample covariance for each $\eta_j$, which requires storage of $J$ $4$ by $4$ matrices. A joint preconditioned MALA update is used for $\epsilon_{1:J}$ and $\kappa$ with diagonal preconditioning matrix given by the sample variances of the $\epsilon_{1:J}$ and $\kappa$ particles. The covariance matrix for the MALA proposal in PGAS, $S_d$ is chosen as the diagonal matrix of the variances of the current set of $f_{1:J, 1:D}$ particles. All preconditioning matrices are calculated after the resampling step, but before tuning and mutation.
	
	The target acceptance rate for all MALA and random walk stepsizes is $\alpha_{\textrm{target}} = 0.6$. Stepsizes are tuned by running the mutation algorithm, i.e.\ MALA or PGAS (without actually mutating the particles), calculating the acceptance rate (or the average movement in PGAS as defined in \eqref{eqn:pgas_ar}) and setting $\epsilon = \epsilon \times \exp{(\alpha - \alpha_{\textrm{target}})}$. Once the acceptance rate is within $0.1$ of $\alpha_{\textrm{target}}$, tuning is stopped. The maximum number of tuning iterations is $10$. Tuning is done after resampling on the unique particles only.
	
	To assess the performance of our SMC$^2$ method, we use two types of cross-validation. The first is blocked leave-future-out cross-validation (LFO-CV), where the model is trained on data up to a specific year, and then predicted on all points in that year \citep{roberts2017,burkner2020}. The second is $10$-fold cross-validation, following \citet{foo2024}, where the inference is run 10 times, each time with a different 10\% of the data withheld as the test set. 
	
	For all runs, we compare the mean error (ME) and mean absolute error (MAE) between the median of the predicted prevalences and the observed prevalences. We also use the non-randomised probability integral transform (PIT) approach of \citet{czado2009} to assess model calibration. The main idea of PIT for continuous data, is that if the observations are drawn from the predictive distribution, then the PIT values, i.e.\ the predictive cumulative distribution function (CDF) evaluated at the observations, have a standard uniform distribution. For count data, the PIT values are no longer uniform as the predictive distribution is discrete. In this case, randomised PIT or the non-randomised version of \citet{czado2009} can be used. We use a qq plot to compare the non-randomised PIT values of \citet{czado2009} to the theoretical quantiles of the standard uniform distribution.

	\subsection{Three Marker Synthetic Data}
	
	To get a baseline for the possible predictive performance of the model given the temporal and spatial coverage of the available data, we first consider a synthetic dataset simulated using the same covariates $x_d$, $d = 1,\ldots, 234$ as in the real 3 marker dataset. To reduce uncertainty due to missing data and low sample sizes, observations are drawn for all $8$ haplotypes and $234$ locations, with all sample sizes set to $1000$, i.e.\ $n_{i,d} = 1000$ for $i = 1,\ldots, 8$, $d = 1, \ldots, 234$. 
	
	Tables \ref{table:dhps3_syn_params1}--\ref{table:dhps3_syn_params3} in Appendix \ref{A:example_syn} show the true values used to simulate the data, as well as the posterior means and standard deviations of $\phi$, $\eta_{1:J}$, $\epsilon_{1:J}$ and $\kappa$. The NNGP covariance parameters $\eta_{1:J}$, and the likelihood dispersion parameter $\kappa$, are all underestimated. This may be due to weak identifiability of the GP covariance parameters \citep{zhang2004}, as well as confounding between the noise parameters of the NNGPs and likelihood. 
	
\begin{figure}[htp]
		\centering
		\includegraphics[scale=0.8]{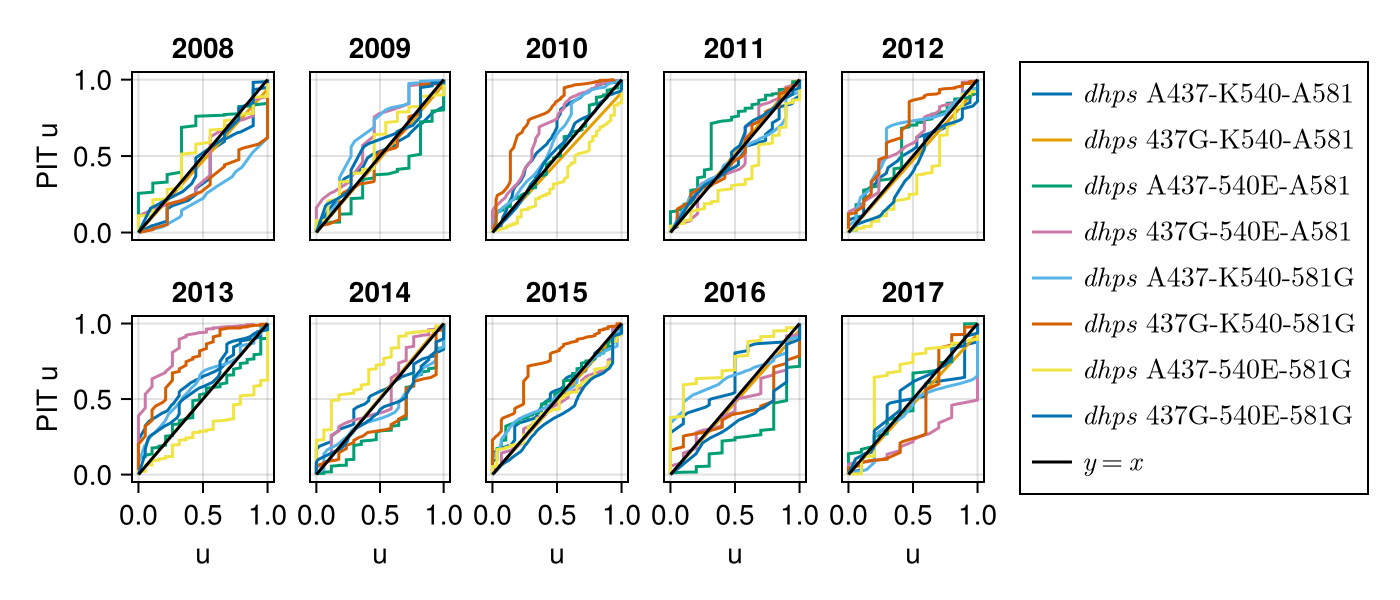}
		\caption{Randomised PIT plots for LFO-CV for the synthetic 3 marker dataset. The number of test points for years 2008 to 2017 are 9, 11, 36, 19, 17, 19, 17, 29, 10 and 10 respectively. The line $y=x$ shows the $U(0,1)$ reference.}
		\label{fig:dhps3_syn_lfocv_pit}
	\end{figure}

\begin{figure}[htp]
		\centering
		\includegraphics[scale=0.8]{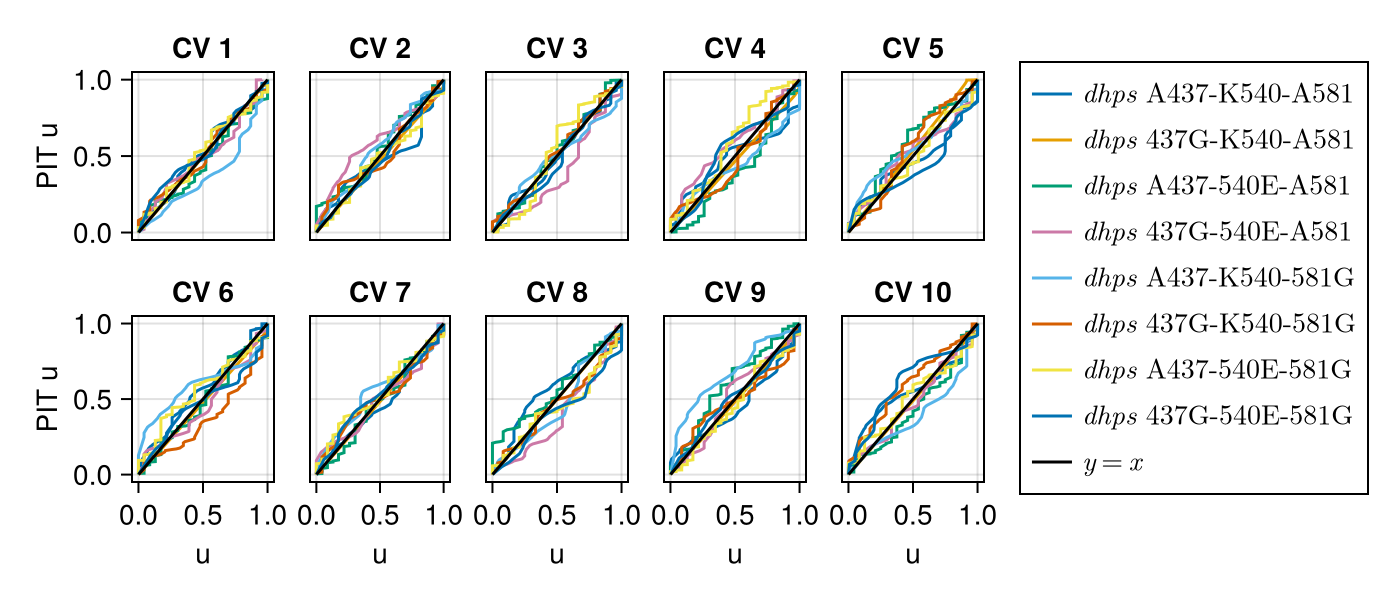}
		\caption{Randomised PIT plots for 10-fold CV for the synthetic 3 marker dataset. The sample sizes are 23 or 24 for all folds. The line $y=x$ shows the $U(0,1)$ reference.}
		\label{fig:dhps3_syn_kfcv_pit}
	\end{figure}

	Table \ref{table:dhps3_syn_error} shows the ME and MAE between the predictive median and the observations for the two CV approaches. Despite the bias in the other parameters, the ME and MAE shows good agreement between the predicted and observed haplotype prevalences. For \textit{dhps} A437-540E-A581 and \textit{dhps} A437-540E-581G the MAE is around 3.5\% for 10-fold CV and 5\% for LFO-CV, but for the remaining haplotypes the ME and MAE are relatively small. Figures \ref{fig:dhps3_syn_lfocv_pit} and \ref{fig:dhps3_syn_kfcv_pit} show that the PIT calibration plots are closely centered around the diagonal line across all haplotypes for 10-fold CV, but are considerably wider for LFO-CV, indicating bias for some of the haplotypes. Note that this is not reflected in Table \ref{table:dhps3_syn_error} as the latter averages the ME and MAE over the years. These results indicate that the available data may provide insufficient support for temporal extrapolation to future years, but that spatial interpolation to missing areas within the same or previous years may be feasible.
       
	Table \ref{table:dhps3_syn_compute} shows the computation time, number of SMC$^2$ iterations and number of datapoints per CV run. While the number of SMC$^2$ iterations increases with the number of datapoints, the computation time per iteration and datapoint is mostly constant across all runs, with some variability. The variability in the run times is likely due to the adaptive nature of the algorithm, e.g.\ longer runs may reflect an earlier switch to a random walk or MALA proposal in PGAS. Results are obtained within 4 hours for all runs, and the median time for the 10-fold CV runs is approximately 2.9 hours.

	\begin{table}[htp]
		\centering
		\begin{tabular}{l|rrr|rrr}
			\hline
			& \multicolumn{3}{c|}{LFO-CV} & \multicolumn{3}{c}{10-fold CV} \\
			Observed haplotype & $n$ & ME & MAE & $n$ & ME & MAE \\
			\hline
            \textit{dhps} A437-K540-A581 & 177 & -0.0018 & 0.0054 & 234 & -0.0002 & 0.0047 \\
            \textit{dhps} 437G-K540-A581 & 177 & 0.0001 & 0.0001 & 234 & 0.0001 & 0.0001 \\
            \textit{dhps} A437-540E-A581 & 177 & -0.0049 & 0.0409 & 234 & -0.0056 & 0.0344 \\
            \textit{dhps} 437G-540E-A581 & 177 & -0.0038 & 0.0095 & 234 & -0.0007 & 0.0071 \\
            \textit{dhps} A437-K540-581G & 177 & -0.0008 & 0.0037 & 234 & 0.0000 & 0.0035 \\
            \textit{dhps} 437G-K540-581G & 177 & -0.0064 & 0.0122 & 234 & -0.0008 & 0.0087 \\
            \textit{dhps} A437-540E-581G & 177 & 0.0115 & 0.0456 & 234 & 0.0025 & 0.0349 \\
            \textit{dhps} 437G-540E-581G & 177 & 0.0002 & 0.0030 & 234 & 0.0003 & 0.0029 \\
			\hline
		\end{tabular}
		\caption{Number of observations, mean error (ME) and mean absolute error (MAE) of the median of the predicted prevalences and the observed prevalences for the synthetic 3 marker dataset. Results are averaged over all test sets in LFO-CV and 10-fold CV.}
		\label{table:dhps3_syn_error}
	\end{table}

	\begin{table}[htp]
		\centering
		\begin{tabular}{ccccc|ccccc}
			\hline
			\multicolumn{5}{c|}{LFO-CV} & \multicolumn{5}{c}{10-fold CV} \\
			Year & Time (hours) & $I$ & $D$ & $s\slash ID$ & CV & Time (hours) & $I$ & $D$ & $s\slash ID$ \\
			\hline
            2008 & 0.44 & 33 & 56 & 0.85 & 1 & 2.20 & 52 & 211 & 0.72  \\
            2009 & 0.53 & 35 & 65 & 0.84 & 2 & 2.37 & 53 & 211 & 0.76  \\
            2010 & 0.81 & 37 & 76 & 1.03 & 3 & 3.06 & 52 & 210 & 1.01  \\
            2011 & 1.34 & 41 & 112 & 1.05 & 4 & 3.93 & 53 & 211 & 1.26  \\
            2012 & 1.75 & 44 & 131 & 1.09 & 5 & 3.67 & 52 & 210 & 1.21  \\
            2013 & 2.15 & 48 & 148 & 1.09 & 6 & 2.65 & 51 & 211 & 0.89  \\
            2014 & 1.59 & 46 & 167 & 0.75 & 7 & 2.09 & 53 & 211 & 0.67  \\
            2015 & 2.35 & 49 & 184 & 0.94 & 8 & 3.24 & 54 & 210 & 1.03  \\
            2016 & 3.16 & 50 & 213 & 1.07 & 9 & 3.81 & 55 & 211 & 1.18  \\
            2017 & 2.40 & 55 & 223 & 0.70 & 10 & 2.64 & 50 & 210 & 0.91  \\
			\hline
		\end{tabular}
		\caption{Total computation time in hours, number of SMC$^2$ iterations ($I$), number of observations in the training set ($D$) and the time in seconds per iteration and datapoint ($s\slash ID$) for the synthetic 3 marker dataset. Results are shown for each of the LFO-CV and 10-fold CV runs. Year refers to the prediction year, e.g.\ the value 2010 means the model was trained on all data before 2010 and tested on all data from 2010.}
		\label{table:dhps3_syn_compute}
	\end{table}

	\subsection{Three Marker Real Data}
	
	We now fit the model to the real 3 \textit{dhps} marker dataset. Here, the sample sizes are smaller, ranging from 3-1259 with a median sample size of 77, and the studies report on a subset of the observed haplotypes shown in Table \ref{table:dhps3_real_error}.
	
\begin{figure}[htp]
		\centering
		\includegraphics[scale=0.8]{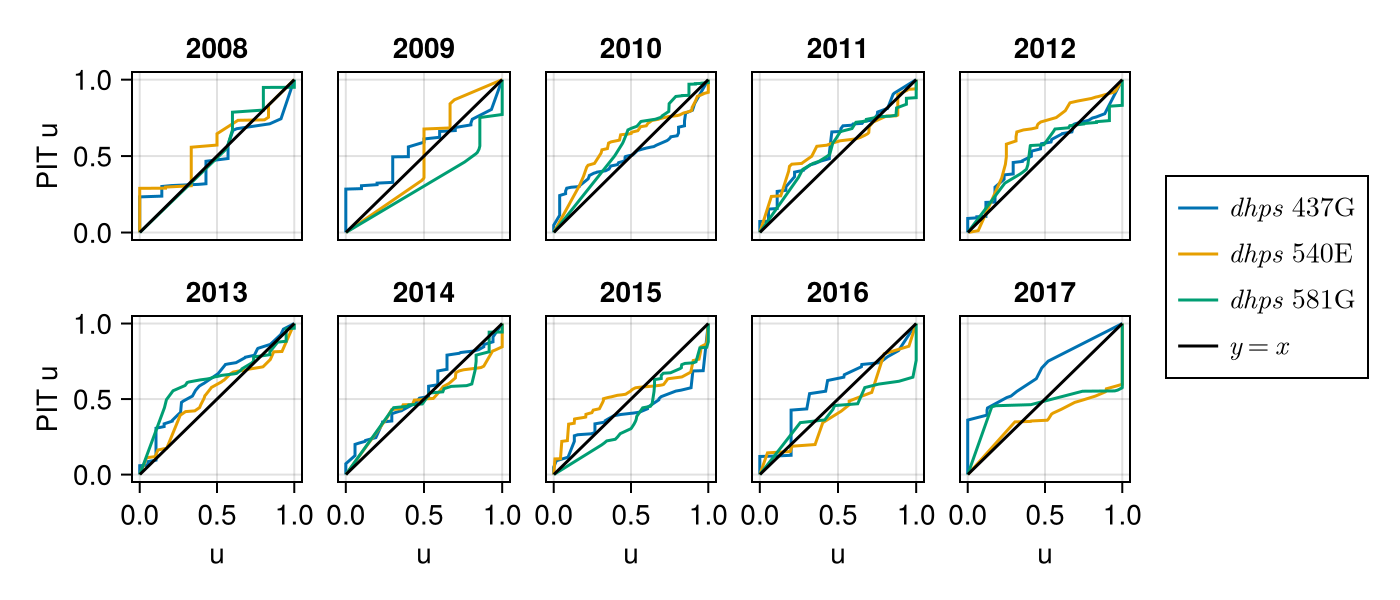}
		\caption{Randomised PIT plots for LFO-CV for the real 3 marker dataset. The number of test points for years 2008 to 2017 are 9, 11, 36, 19, 17, 19, 17, 29, 10 and 10 respectively. The line $y=x$ shows the $U(0,1)$ reference.}
		\label{fig:dhps3_real_lfocv_pit}
	\end{figure}

\begin{figure}[htp]
		\centering
		\includegraphics[scale=0.8]{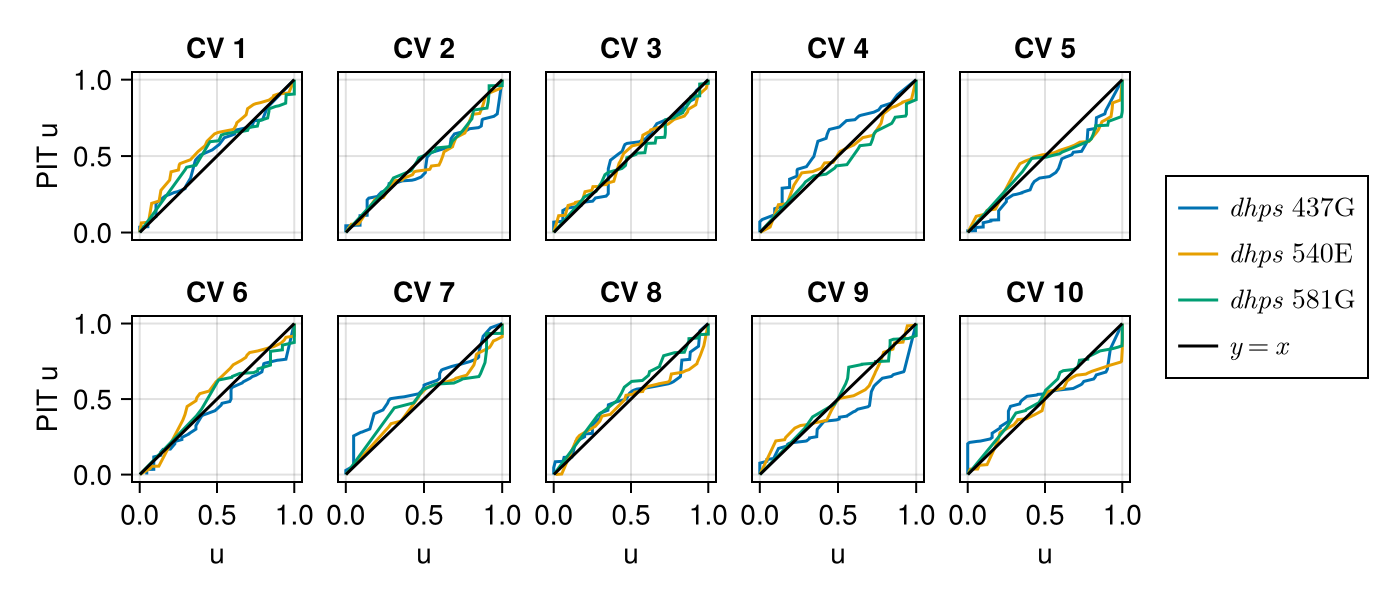}
		\caption{Randomised PIT plots for 10-fold CV for the real 3 marker dataset. The number of test points is $23$ or $24$ for all folds. The line $y=x$ shows the $U(0,1)$ reference.}
		\label{fig:dhps3_real_kfcv_pit}
	\end{figure}

	Table \ref{table:dhps3_real_error} shows the number of observations for each observed haplotype, as well as the ME and MAE for the LFO-CV and 10-fold CV runs. The MAE for 10-fold CV is similar to the results obtained by \citet{foo2024}, but the ME is slightly worse in general. The latter may be due to the approximate model, or simply randomness in the results. The ME and MAE are consistently worse for LFO-CV than for 10-fold CV, which is unsurprising given the synthetic data results. Similarly, Figures \ref{fig:dhps3_real_lfocv_pit} and \ref{fig:dhps3_real_kfcv_pit} show reasonable PIT calibration for 10-fold CV but poor results for LFO-CV.
	
	Table \ref{table:dhps3_real_compute} shows the computation times for the real data. Again, the cost of each iteration of SMC$^2$ is roughly linear in the number of datapoints, and results are obtained within 4 hours for all runs. The median run time for the 10-fold CV runs is approximately 2.7 hours.
	
	\begin{table}[htp]
		\centering
		\begin{tabular}{l|rrr|rrr}
			\hline
			& \multicolumn{3}{c|}{LFO-CV} & \multicolumn{3}{c}{10-fold CV} \\
			Observed haplotype & $n$ & ME & MAE & $n$ & ME & MAE \\
			\hline
            \textit{dhps} 437G & 147 & 0.0118 & 0.1306 & 202 & 0.0421 & 0.1177 \\
            \textit{dhps} 540E & 156 & -0.0320 & 0.1045 & 207 & 0.0134 & 0.0739 \\
            \textit{dhps} 581G & 130 & -0.0507 & 0.0786 & 148 & -0.0446 & 0.0628 \\
            \textit{dhps} 437G-540E & 34 & -0.0750 & 0.1244 & 64 & 0.0132 & 0.0549 \\
            \textit{dhps} 437G-540E-581G & 37 & 0.0055 & 0.0827 & 40 & 0.0239 & 0.0706 \\
            \textit{dhps} 437G-540E-A581 & 39 & -0.0758 & 0.0793 & 42 & -0.0572 & 0.0687 \\
			\hline
		\end{tabular}
		\caption{Number of observations, mean error (ME) and mean absolute error (MAE) of the median of the predicted prevalences and the observed prevalences for the real 3 marker dataset. Results are averaged over all test sets in LFO-CV and 10-fold CV.}
		\label{table:dhps3_real_error}
	\end{table}
	
	\begin{table}[htp]
		\centering
		\begin{tabular}{ccccc|ccccc}
			\hline
			\multicolumn{5}{c|}{LFO-CV} & \multicolumn{5}{c}{10-fold CV} \\
			Year & Time (hours) & $I$ & $D$ & $s\slash ID$ & CV & Time (hours) & $I$ & $D$ & $s\slash ID$ \\
			\hline
            2008 & 0.27 & 31 & 56 & 0.56 & 1 & 2.65 & 47 & 211 & 0.96  \\
            2009 & 0.37 & 32 & 65 & 0.65 & 2 & 2.25 & 48 & 211 & 0.80  \\
            2010 & 0.61 & 33 & 76 & 0.87 & 3 & 3.11 & 51 & 210 & 1.04  \\
            2011 & 0.62 & 36 & 112 & 0.55 & 4 & 3.01 & 48 & 211 & 1.07  \\
            2012 & 1.55 & 39 & 131 & 1.09 & 5 & 2.81 & 48 & 210 & 1.00  \\
            2013 & 1.60 & 42 & 148 & 0.93 & 6 & 2.90 & 49 & 211 & 1.01  \\
            2014 & 2.06 & 43 & 167 & 1.03 & 7 & 1.86 & 51 & 211 & 0.62  \\
            2015 & 2.12 & 44 & 184 & 0.94 & 8 & 2.29 & 48 & 210 & 0.82  \\
            2016 & 2.81 & 48 & 213 & 0.99 & 9 & 3.35 & 51 & 211 & 1.12  \\
            2017 & 3.31 & 49 & 223 & 1.09 & 10 & 2.21 & 48 & 210 & 0.79  \\
			\hline
		\end{tabular}
		\caption{Total computation time in hours, number of SMC$^2$ iterations ($I$), number of observations in the training set ($D$) and the time in seconds per iteration and datapoint ($s\slash ID$) for the real 3 marker dataset. Results are shown for each of the LFO-CV and 10-fold CV runs. Year refers to the prediction year, e.g.\ the value 2010 means the model was trained on all data before 2010 and tested on all data from 2010.}
		\label{table:dhps3_real_compute}
	\end{table}

	\subsection{Six Marker Real Data}
	
\begin{figure}[htp]
		\centering
		\includegraphics[scale=0.8]{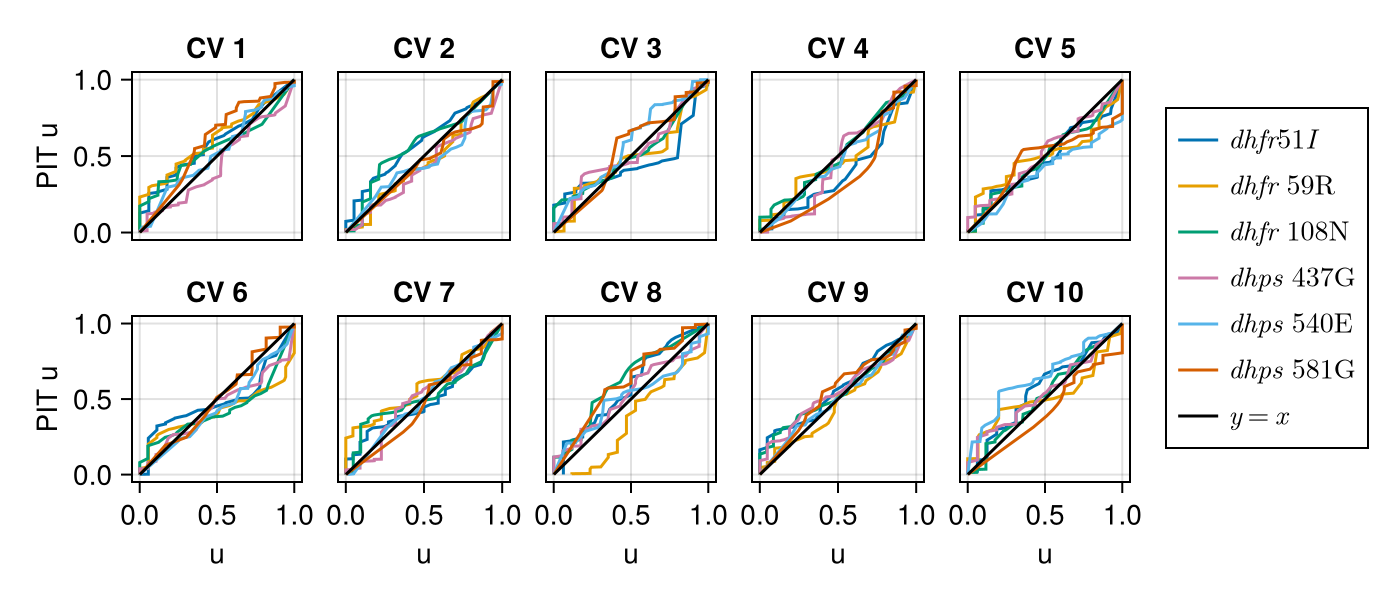}
		\caption{Randomised PIT plots for 10-fold CV for the 6 marker dataset. The number of test points is $24$ or $25$ for all folds. The line $y=x$ shows the $U(0,1)$ reference.}
		\label{fig:dhps3dhfr3_real_kfcv_pit}
	\end{figure}

	Our final example fits the model to the 6 marker dataset, comprising the 3 \textit{dhps} and 3 \textit{dhfr} markers. The observed haplotypes are shown in Table \ref{table:dhps3dhfr_real_error}. Considering the poor results for LFO-CV in the 3 marker datasets, we only run 10-fold CV for this example. Table \ref{table:dhps3dhfr_real_error} shows the ME and MAE results, and Table \ref{table:dhps3dhfr_real_compute} shows the computation times. The ME and MAE show clear bias for some of the observed haplotypes, with the MAE ranging from 3\% to 20\%. This may be due to insufficient multimarker observations for the number of haplotype prevalences being estimated. The PIT plots in Figure \ref{fig:dhps3dhfr3_real_kfcv_pit} shows reasonable calibration for most of the folds, however, there is some indication of model misspecification in folds 5 to 7.
	
	Table \ref{table:dhps3dhfr_real_compute} shows a roughly linear computation cost for each iteration of SMC$^2$ in the number of datapoints and the number of NNGPs when compared to the 3 marker results. Results are obtained within 43 hours for all runs, and the median run time is 23.2 hours (an 8.6 fold increase from the median time for the 3 marker data). 
		
	\begin{table}[htp]
		\centering
		\begin{tabular}{lrrr}
			\hline
			Observed haplotype & $n$ & ME & MAE \\
			\hline
            \textit{dhps} 540E & 201 & -0.0305 & 0.1327 \\
            \textit{dhps} 437G & 193 & 0.0106 & 0.1597 \\
            \textit{dhfr} 108N & 183 & -0.0017 & 0.0949 \\
            \textit{dhfr} 59R & 177 & 0.0120 & 0.1488 \\
            \textit{dhfr} 51I & 176 & 0.0002 & 0.1259 \\
            \textit{dhps} 581G & 145 & -0.0119 & 0.0878 \\
            \textit{dhfr} 51I-59R-108N & 120 & 0.0030 & 0.1879 \\
            \textit{dhps} 437G-540E & 60 & -0.0268 & 0.1159 \\
            \textit{dhfr}+\textit{dhps} 51I-59R-108N-437G-540E & 57 & 0.0042 & 0.0907 \\
            \textit{dhfr} N51-59R-108N & 48 & -0.0233 & 0.0868 \\
            \textit{dhps} 437G-540E-581G & 41 & -0.0566 & 0.0843 \\
            \textit{dhps} 437G-540E-A581 & 39 & 0.0088 & 0.0942 \\
            \textit{dhfr}+\textit{dhps} 51I-59R-108N-437G-540E-581G & 19 & -0.0168 & 0.0352 \\
            \textit{dhfr}+\textit{dhps} 51I-59R-108N-437G-540E-A581 & 14 & 0.0710 & 0.0710 \\
			\hline
		\end{tabular}
		\caption{Number of observations, mean error (ME) and mean absolute error (MAE) of the median of the predicted prevalences and the observed prevalences for the 6 marker dataset. Results are averaged over all test sets in 10-fold CV.}
		\label{table:dhps3dhfr_real_error}
	\end{table}
	
	\begin{table}[htp]
		\centering
		\begin{tabular}{ccccc}
			\hline
			CV & Time (hours) & $I$ & $D$ & $s\slash ID$ \\
			\hline
            1 & 39.82 & 44 & 219 & 14.88  \\
            2 & 22.14 & 49 & 219 & 7.43  \\
            3 & 19.20 & 44 & 219 & 7.17  \\
            4 & 21.98 & 48 & 218 & 7.56  \\
            5 & 20.21 & 47 & 219 & 7.07  \\
            6 & 24.22 & 48 & 219 & 8.29  \\
            7 & 26.19 & 51 & 218 & 8.48  \\
            8 & 42.51 & 48 & 219 & 14.56  \\
            9 & 19.95 & 46 & 219 & 7.13  \\
            10 & 39.18 & 44 & 218 & 14.70  \\
			\hline
		\end{tabular}
		\caption{Total computation time in hours, number of SMC$^2$ iterations ($I$), number of observations in the train set ($D$) and the time in seconds per iteration and datapoint ($s\slash ID$) for the 6 marker dataset. Results are shown for each of the 10-fold CV runs.}
		\label{table:dhps3dhfr_real_compute}
	\end{table}
	
	\section{Discussion} \label{sec:discussion}
	
	In this paper, we develop an approximation to the latent multinomial model of \citet{foo2024} for pooled genetic data. Specifically, we make two important changes, both of which significantly reduce the computational complexity of the model. First, we replace the full GPs with NNGPs, reducing the complexity from $O(JD^3)$ to $O(JDk^3)$, where $J$ is the number of GPs\slash NNGPs, $D$ is the number of observations and $k$ is the number of neighbors. This has the additional benefit of reducing the computational cost of predicting over a large number of new locations, e.g.\ when generating predictive maps as in \citet{flegg2024} and \citet{foo2024}. Second, we replace the multinomial likelihood with a composite marginal likelihood, where the weights are chosen to reduce redundancy. Unlike the multinomial likelihood, the composite likelihood does not require the latent counts to be inferred, thereby reducing the computational burden and allowing for measurement error, and also accommodates different sample sizes within the same study. 
	
	To infer the parameters of the model, we develop a new SMC$^2$ method which uses PGAS to mutate the NNGP function values. Our PGAS algorithm uses a combination of three proposal distributions, including a bootstrap, random walk and MALA proposal. The latter especially is designed to scale favorably with both the number of observations and the dimension of the state vector (i.e.\ the number of NNGPs) \citep{corenflos2024}. Our approach is fully automatic, and we show that the computational complexity of a single PGAS update has a linear cost in the number of observations and the number of NNGPs, i.e.\ the computational cost is $O(JDk^3)$ (see Section \ref{sssec:cost}). This is also shown empirically in our examples in Section \ref{sec:example}, as the computation time per SMC$^2$ iteration is roughly linear in $D$ and $J$, with some variation due to the adaptive nature of the algorithm. 

    Our approach can be seen as a general inference method for the NNGP function values. Alternative approaches rely on Gibbs or Metropolis-within-Gibbs updates \citep{datta2016,datta2016a}, where the NNGP function values are updated sequentially by location or in blocks. While sequential updates are computationally cheaper than joint updates, they may lead to mixing problems in high dimensions \citep{coube-sisqueille2022}. Hamiltonian Monte Carlo \citep[HMC; ][]{neal2011} is an option for joint updates and have been used for GP models \citep{foo2024}, but each HMC transition requires $L$ leapfrog steps. If the cost of each step is $O(JDk^3)$, then the cost per HMC transition is $O(LJDk^3)$. \citet{beskos2013} show that, in a high-dimensional setting where the target factorises into $d$ independent and identically distributed components, the stepsize should scale as $d^{-1\slash 4}$ to keep a non-degenerate acceptance rate. For a fixed length trajectory, this implies $L = O(d^{1\slash 4})$, where $d = JD$ in our setting. Then, for a joint update of the latent values from $J$ NNGPs over $D$ observations, the heuristic cost of a single HMC transition is $O(LJDk^3) = O((JD)^{5\slash 4}k^3)$ compared to $O(JDk^3)$ as in our approach.

    In our examples, we used the same spatio-temporal NNGP model, but our method is straightforward to apply to a broad range of NNGP models, e.g.\ NNGP extensions of state-space GP models \citep{frigola2013} or latent factor NNGP models \citep{taylor-rodriguez2019,davies2022}. More work is needed to extend our approach to dependent NNGPs, where cross-process dependence is encoded directly at the level of the NNGP function values \citep{datta2016,grenier2024}, rather than induced by combining independent NNGPs through a loading or mixing structure. Dependency at this level will induce a cubic cost in $J$ due to the MALA proposal (see Section \ref{sssec:cost}). 
	
	As a case study, we consider molecular marker data relating to antimalarial drug resistance in Africa. We fit our model to a dataset with 3 \textit{dhps} markers, as well as 3 \textit{dhps} and 3 \textit{dhfr} markers. \citet{foo2024} also inferred the prevalences of the 3 marker \textit{dhps} haplotypes, but computational limitations prevent their approach from scaling to a larger number of markers. Earlier work \citep{flegg2013,flegg2022} only considered single marker mutations. To date, Bayesian inference of haplotype prevalences has not been done for 6 markers, or for multi-gene haplotypes. We investigate the spatio-temporal coverage of the available data by fitting the model to synthetic data generated using the same covariates as in the 3 marker dataset. To reduce uncertainty, we assume full observation of the haplotypes and a sample size of $1000$ for all counts. We find that care should be taken when interpreting any of the NNGP and likelihood parameters, as they are only weakly identifiable for our model. Based on this example, the available data are too sparse to provide reliable predictions for future years. This is exacerbated when full observation of the haplotypes are not available, and likely also by changes in reporting over the years, e.g.\ few of the earlier studies report on the \textit{dhps} A581G marker. Despite this, interpolation of unseen locations for current or previous years seems feasible for a smaller number of markers. There is good agreement between the predictive and observed haplotype prevalances for the 10-fold cross-validation sets for both of the 3 marker datasets. 
	
	For the 6 marker dataset, some of the haplotype prevalences show clear bias. This may be due to insufficient data, as there are few multimarker observations in the 6 marker dataset relative to the number of haplotypes being estimated. A simulation study would be useful to identify potential issues with the data for estimating a large number of haplotype prevalences. Predictive performance may also be improved by considering a different model, e.g.\ a latent factor model with $J = 2^{G-1}$ rather than $J = 2^G-1$ NNGPs, or by including additional covariates in the model, such as the percentage of IPTp use by SP dose \citep{who2025}. This is left as future work. 

    Our model could be used to undertake analyses to support more strategic use of SP as an IPTp in future. The current WHO recommendation is for at least 3 doses of SP to be administered to all pregnant women in malaria-endemic areas in Africa, starting in the second trimester \cite{who2025}. A WHO Evidence Review Group \citep{who2013} previously proposed that SP be discontinued as an IPTp in areas where the prevalence of the \textit{dhps} K540E mutation exceeds 95\% and the prevalence of the \textit{dhps} A581G mutation exceeds 10\%. However, this relies on the prevalence of single marker mutations when reduced effectiveness of SP as an IPTp has been explicitly linked to the sextuple mutant haplotype involving the 3 \textit{dhps} and 3 \textit{dhfr} markers \citet{vaneijk2019}. The current work allows the relevant sextuple mutation to be modelled directly, which may better inform decisions about the use of SP as an IPTp in areas of high malaria transmission.

	\section*{Acknowledgements}
	This research was supported by The University of Melbourne’s Research Computing Services and the Petascale Campus Initiative. J.A.\ Flegg’s research was supported by the Australian Research Council (FT210100034, CE230100001) and the National Health and Medical Research Council (APP2019093). N. Golding's research was supported by the Stan Perron Charitable Foundation and an NHMRC Investigator Grant (2041810).  

	\bibliographystyle{apalike} 
	\bibliography{refs}
	
	\appendix
	
	\section{Supplementary Results for the Three Marker Synthetic Data Example} \label{A:example_syn}
	
	\begin{table}[H]
		\centering
		\begin{tabular}{|l|cc|cc|cc|}
			\hline
			j  & \multicolumn{2}{c|}{$\mu_j$} & \multicolumn{2}{c|}{$\beta_j$} & \multicolumn{2}{c|}{$s_j$} \\
			\hline
			1 & 4.15, & 4.08 (0.19) & -0.35, & -1.87 (0.56) & 0.15, & 0.17 (0.07) \\
            2 & -2.85, & -3.76 (0.14) & -2.43, & -1.53 (0.30) & 0.60, & 0.44 (0.07) \\
            3 & -1.13, & -1.31 (0.09) & 0.34, & 1.10 (0.26) & 0.55, & 0.32 (0.08) \\
            4 & 0.32, & 0.22 (0.10) & -1.96, & -1.43 (0.29) & 0.24, & 0.18 (0.06) \\
            5 & -0.53, & -0.83 (0.07) & -2.78, & -2.05 (0.19) & 0.21, & 0.16 (0.04) \\
            6 & -5.25, & -4.53 (0.09) & 0.97, & 1.15 (0.14) & 0.59, & 0.32 (0.04) \\
            7 & 1.64, & 1.52 (0.09) & -2.28, & -2.10 (0.18) & 0.32, & 0.16 (0.05) \\
			\hline
		\end{tabular}
		\caption{True values, posterior means and posterior standard deviations (true value, posterior mean (posterior standard deviation)) for the parameters $\mu_{1:J}$, $\beta_{1:J}$ and $s_{1:J}$ for the synthetic 3 marker dataset.}
		\label{table:dhps3_syn_params1}
	\end{table}
	
	\begin{table}[H]
		\centering
		\begin{tabular}{|l|cc|cc|cc|}
			\hline
			j  & \multicolumn{2}{c|}{$\sigma_j$} & \multicolumn{2}{c|}{$\tau_j$} & \multicolumn{2}{c|}{$\delta_j$} \\
			\hline
            1 & 0.28, & 0.18 (0.05) & 1.03, & 0.56 (0.15) & 1.21, & 0.83 (0.32) \\
            2 & 0.16, & 0.17 (0.04) & 0.82, & 0.55 (0.14) & 1.89, & 0.81 (0.31) \\
            3 & 0.21, & 0.19 (0.05) & 0.92, & 0.55 (0.14) & 1.37, & 0.83 (0.31) \\
            4 & 0.22, & 0.18 (0.04) & 0.94, & 0.55 (0.15) & 1.16, & 0.83 (0.31) \\
            5 & 0.16, & 0.15 (0.03) & 0.94, & 0.56 (0.15) & 1.68, & 0.83 (0.32) \\
            6 & 0.13, & 0.11 (0.02) & 1.17, & 0.64 (0.15) & 1.91, & 0.80 (0.28) \\
            7 & 0.14, & 0.17 (0.04) & 1.17, & 0.55 (0.15) & 1.49, & 0.84 (0.33) \\
			\hline
		\end{tabular}
		\caption{True values, posterior means and posterior standard deviations (true value, posterior mean (posterior standard deviation)) for the parameters $\sigma_{1:J}$, $\tau_{1:J}$ and $\delta_{1:J}$ for the synthetic 3 marker dataset.}
		\label{table:dhps3_syn_params2}
	\end{table}
	
	\begin{table}[H]
		\centering
		\begin{tabular}{|cc|cc|cc|}
			\hline
			\multicolumn{2}{|c|}{$\kappa$} & \multicolumn{2}{c|}{$\beta_{s}$} & \multicolumn{2}{c|}{$\beta_{\sigma}$} \\
			\hline
			772.30, & 311.37 (11.83) & 1.38, & 1.09 (0.25) & 0.83, & 0.78 (0.16) \\
			\hline
		\end{tabular}
		\begin{tabular}{|cc|cc|}
			\hline
			\multicolumn{2}{|c|}{$\beta_{\tau}$} & \multicolumn{2}{c|}{$\beta_{\delta}$} \\
			\hline
			47.02, & 27.72 (6.35) & 77.32, & 40.33 (14.10) \\
			\hline
		\end{tabular}
		\caption{True values, posterior means and posterior standard deviations (true value, posterior mean (posterior standard deviation)) for the parameters $\kappa$ and $\phi$ for the synthetic 3 marker dataset.}
		\label{table:dhps3_syn_params3}
	\end{table}

\end{document}